\documentclass[12pt,tabularx,amsmath,a4paper]{article}

\usepackage{setspace,graphics}
\usepackage[dvips]{epsfig} 
\usepackage{a4,amssymb,epsfig,array,cite,setspace,amsmath}
\usepackage[hypertex]{hyperref}
\usepackage{slashed}

\newcounter{multieqs}

\newcommand{\be}{\begin{equation}}
\newcommand{\ee}{\end{equation}}
\newcommand{\eq}[1]{(\ref{#1})}

\def\nn{\nonumber}
\def\bea{\begin{eqnarray}}
\def\eea{\end{eqnarray}}

\def\beqa{\begin{eqnarray}} 
\def\eeqa{\end{eqnarray}} 
\def\beq{\begin{equation}} 
\def\eeq{\end{equation}}

\def\Tr{{\rm Tr}}

 \def\L{\Lambda}

\def\cA{{\cal A}}  \def\cC{{\cal C}}
 \def\cE{{\cal E}} 
 \def\cH{{\cal H}} 
  
\def\cM{{\cal M}}  \def\cO{{\cal O}}

\def\R{{\mathbb R}}

\def\one{\mbox{1 \kern-.59em {\rm l}}}

\def\bit{\begin{itemize}}
\def\eit{\end{itemize}}

\def\({\left(}
\def\){\right)}

\def\uno{\mbox{1 \kern-.59em {\rm l}}}

\newcommand{\tr}{\mbox{tr}}

\def\bcomment#1{}

 \def\ii{{\rm i\,}}

\newcommand{\umu}{_{\mu}}
\newcommand{\omu}{^{\mu}}
\newcommand{\unu}{_{\nu}}
\newcommand{\onu}{^{\nu}}
\newcommand{\ualpha}{_{\alpha}}

\newcommand{\ubeta}{_{\beta}}
\newcommand{\obeta}{^{\beta}}
\newcommand{\umunu}{_{\mu\nu}}
\newcommand{\omunu}{^{\mu\nu}}
\newcommand{\ualphabeta}{_{\alpha\beta}}
\newcommand{\oalphabeta}{^{\alpha\beta}}
\newcommand{\urho}{_{\rho}}
\newcommand{\orho}{^{\rho}}
\newcommand{\usigma}{_{\sigma}}
\newcommand{\osigma}{^{\sigma}}
\newcommand{\orhosigma}{^{\rho\sigma}}
\newcommand{\urhosigma}{_{\rho\sigma}}
\newcommand{\ulambda}{_{\lambda}}
\newcommand{\olambda}{^{\lambda}}
\newcommand{\ugamma}{_{\gamma}}

\newcommand{\udelta}{_{\delta}}
\newcommand{\odelta}{^{\delta}}
\newcommand{\oprime}{^{\prime}}

\newcommand{\bsp}{\begin{split}}
\newcommand{\esp}{\end{split}}

\newcommand{\thetainv}{\theta^{-1}}
\newcommand{\Gtilde}{\widetilde{G}}
\newcommand{\ems}{e^{-\sigma}}
\newcommand{\eps}{e^{\sigma}}
\newcommand{\hme}{^{-1}}
\newcommand{\ad}{\mathrm{d}}
\newcommand{\gammat}{\widetilde{\gamma}}

\renewcommand{\title}[1]{\vspace{10mm}\noindent{\Large{\bf #1}}\vspace{8mm}}
\newcommand{\authors}[1]{\noindent{\large #1}\vspace{5mm}}
\newcommand{\address}[1]{{\itshape #1\vspace{2mm}}}

\begin{document}
\bibliographystyle{JHEP}
\begin{titlepage}
\begin{flushright}
UWTHPh-2009-09\\
\end{flushright}

\begin{center}
  
\title{Fermions and noncommutative emergent
  gravity II:\\\vspace{3mm} Curved branes in extra dimensions} \\
\vspace{2cm}

\authors{Daniela {\sc Klammer}${}^{1}$ and Harold {\sc Steinacker}${}^{2}$}

\address{ Fakult\"at f\"ur Physik, Universit\"at Wien\\
 Boltzmanngasse 5, A-1090 Wien, Austria \\
 \today}

\footnotetext[1]{daniela.klammer@univie.ac.at}
\footnotetext[2]{harold.steinacker@univie.ac.at}

\vskip 2cm

\textbf{Abstract}

\vskip 3mm 

\begin{minipage}{14cm}%

We study fermions coupled to Yang-Mills matrix models from the point of view of
emergent gravity. The matrix model Dirac operator provides
an appropriate coupling for fermions to the effective gravitational metric 
for general branes nontrivial embedding, 
albeit with a non-standard spin connection. This generalizes previous 
results for 4-dimensional matrix models.
Integrating out the fermions in a nontrivial geometrical background
induces indeed the Einstein-Hilbert action of the 
effective metric,
as well as additional terms which couple 
the Poisson tensor to the Riemann tensor, and a dilaton-like term.

\end{minipage}

\end{center}

\end{titlepage}

\setcounter{page}0
\thispagestyle{empty}

\begin{spacing}{.3}
{
\noindent\rule\textwidth{.1pt}            
   \tableofcontents
\vspace{.6cm}
\noindent\rule\textwidth{.1pt}
}
\end{spacing}

\section{Introduction}

Quantum field theory (QFT) and general relativity (GR) 
provide the basis for our present understanding of fundamental physics. However,
these two theories together imply that classical space-time loses its 
meaning in the small. 
It is expected that the conventional concepts of space and time will no
longer hold at the Planck scale and instead some 
kind of quantum structure of space-time should take over in this
regime. One way of describing such a structure 
is obtained by taking a  \emph{noncommutative algebra} for spacetime coordinates. 
The basic idea is that the 
classical space-time $\R^4$ is replaced by a space where the
 coordinate functions $x\omu$ satisfy Heisenberg-like commutation relations,
\begin{align}
[x\omu,x\onu]=\ii \theta\omunu.
\end{align}
The situation is quite 	analogous to quantum mechanics. At the semi-classical level, these commutation relations reduce to a Poisson structure $\theta\omunu(x)$ on space-time. 
One can then write down so-called \emph{noncommutative quantum field theories} which incorporate quantum fluctuations of spacetime coordinates naturally, see~\cite{Szabo:2001kg}. It was conjectured that these fluctuations should then in some way be linked to gravity. In recent years a specific realization of this idea was developed under the name of ``emergent noncommutative gravity'', see ~\cite{Steinacker:2007dq,Grosse:2008xr,Klammer:2008df,Steinacker:2008ri,Steinacker:2008ya,Klammer:2009ku,Steinacker:2009mp}. 

In this series of papers it was understood that matrix models of Yang-Mills type
\begin{align}
S_{\mathrm{YM}}=-\Tr [X^a,X^b][X^{a\oprime},X^{b\oprime}]\eta_{aa\oprime}\eta_{bb\oprime}
\end{align}
as known  from noncommutative (NC) gauge theory and string theory not only describe dynamical NC spaces, in fact they also incorporate gravity. These models contain an effective metric
\begin{align}\label{eq: Gtilde}
\Gtilde\omunu(x)=e^{-\sigma}\theta^{\mu\alpha}(x)\theta^{\nu\beta}(x)g\ualphabeta(x),
\end{align}
which couples to all types of fields. Both the metric responsible for
gravity as well as space-time are not fundamental objects of the
theory, but rather they are determined by the Poisson structure
$\theta(x)$ and the background metric $g\umunu(x)$.

Fermions can be naturally included in the matrix model. 
The fermionic action is similar to the standard action 
for fermions on curved backgrounds coupled to the effective metric
$\Gtilde\omunu$, however with a non-standard 
spin connection. More precisely, the spin connection vanishes 
in the special (but unobservable) matrix coordinates.
This implies that the induced effective action 
due to integrating out the fermions does not quite have the 
standard form in terms of an induced Einstein-Hilbert action.
We computed this induced action in \cite{Klammer:2008df} in the special case of 
branes with flat embedding but general $\theta^{\mu\nu}(x)$,
showing that the expected Einstein-Hilbert action is indeed induced however
with a non-standard factor, along with an extra scalar term.

In this work, we generalize these results to 
 the general case of non-trivially embedded
branes in a higher-dimensional matrix model.
We study the quantization of fermions in the matrix model, and 
compute the effective gravitational action obtained by 
integrating out the fermions. Due to the non-standard coupling 
of the fermions to gravity we cannot apply the standard results for the 
one-loop effective action that can be found in the literature, 
e.g.~\cite{Vassilevich:2003xt,Gilkey:1995mj}. Instead we have to 
evaluate it directly. We are able to cast the effective action
into a covariant geometrical form. It turns out that it contains indeed
expected Einstein-Hilbert 
term  $R[\Gtilde]$,  plus additional terms which 
involve the curvature tensor 
coupled to $\theta^{\mu\nu}$.
Due to technical complications we focus on two special cases:
1) ``on-shell geometries'' as determined by the 
semi-classical equations of motion of the matrix model, 
and 2) the class of geometries where
the effective metric $\Gtilde\omunu$ coincides with the 
induced metric $g^{\mu\nu}$. The latter class seems to be general
enough for a large class of physical situations \cite{Klammer:2009ku,Steinacker:2009mp}.
Our main result is the effective action \eq{eff-action-final}, 
\eq{eff-action-final-2} in these two
cases which have essentially the same structure,
and alternative covariant expressions \eq{eq: covariant E} and 
\eq{eq: covariant E-2} for the novel terms.

This paper is quite technical and involves lengthy computations.
This is necessary because of the non-standard 
Dirac operator, and the composite nature of the effective metric
which involves the Poisson tensor as well as the embedding metric
given in terms of scalar fields. 
Even the demonstration that the induced effective action 
is a well-defined geometrical quantity is non-trivial and
requires a lot of work. In order to make the paper 
as readable as possible we have delegated much of the computations
to the Appendices.

\section{The matrix model in higher dimensions}
As a starting point consider the matrix model
\begin{align}\label{eq: the model}
S_{\mathrm{YM}}=-(2\pi)^n\,\Tr\left(\frac{1}{4} [X^a,X^b][X^{a\oprime},X^{b\oprime}]\eta_{aa\oprime}\eta_{bb\oprime}
+\frac{1}{2} \bar{\Psi} \gamma_a[X^a,\Psi]\right),
\end{align}
where the $X^a$ for $a=1,\hdots,D$ are infinite dimensional hermitian matrices or operators acting on some Hilbert space $\cH$ and
\begin{align}
\eta_{aa\oprime}=\mathrm{diag}(1,1,\hdots,1)\quad\mathrm{or}\quad \eta_{aa\oprime}=\mathrm{diag}(-1,1,\hdots,1)
\end{align}
is an (unphysical) background metric that fixes the signature of the
theory, Euclidean or Minkowski space, respectively. $\gamma_a$
generate the Clifford algebra in $D$ dimensions, and $\Psi$ are spinors
consisting of Grassmann-valued matrices. 
This model can be obtained e.g. as dimensional reduction of large-N
super-Yang-Mills theory to 0 dimensions. A particularly
important case is the IKKT model \cite{Ishibashi:1996xs}, which was first 
proposed in the context of string theory.

\paragraph{Quantization.}
The basic assumption of the present approach is that space-time carries a Poisson structure 
$\{x^\mu,x^\nu\}=\theta^{\mu\nu}(x)$. Space-time is then considered to be the quantization of such a Poisson manifold. 
It is well-known~\cite{Kontsevich:1997vb} that a Poisson manifold ($\cM,\theta\omunu(x)$) can be quantized and  that there exists a quantization map
\begin{align}
\begin{split}
\cC(\cM) &\rightarrow \cA \subset L(\cH) \\
f(x) &\mapsto \widehat{f}(X) \\
\ii \{f,g\} &\mapsto [\widehat{f},\widehat{g}] + O(\theta^2).
\end{split}
\end{align}
$\cC(\cM)$ denotes some space of functions on $\cM$, and $\cA$ is interpreted as quantized algebra of functions on $\cM$. 
The matrices $X^\mu$ are interpreted as quantization of the coordinate function $x^\mu$. Moreover, for the sake of simplicity we will consider only the semi-classical limit of such a quantum space, i.e. we keep only terms linear in $\theta$. Then we have
\begin{align}
[X\omu, X\onu] &\sim \ii \theta\omunu (x) \\
[X\omu, f(X)] &\sim \ii\theta\omunu \frac{\partial}{\partial x\onu}f(x).
\end{align}
The trace is replaced by an integral  where the appropriate density factor is given by the symplectic volume, 
\begin{align}\label{eq: symplectic volume}
(2\pi)^n\, \Tr\, \widehat{f}(X) &\sim \int \ad^{2n} x \;\rho(x) f(x)\\
\rho(x)&=(\det \theta\hme\umunu)^{1/2}.
\end{align}
$\theta\omunu$ is assumed to be non-degenerate and $\det \theta\omunu > 0$.

 
\paragraph{Embedding.}\label{par: embedding}
We want to study $2n$-dimensional NC spaces $\cM^{2n}\subset\R^D $, which we interpret as space-time manifold embedded in $D$ dimensions. To realize this we split the matrices as
\begin{align}
X^a = (X\omu, \phi^i),\qquad \mu=1,\hdots,2n, \quad i=1,\hdots,D-2n,
\end{align}
where the ``scalar fields'' $\phi^i=\phi^i(X\omu)$ are assumed to be functions of $X\omu$ which determine the embedding of a $2n$-dimensional submanifold $\cM^{2n}$ in $\R^D$. $\cM^{2n}$ carries then the induced metric
\begin{align}\label{eq: background metric}
g\umunu (x) = \eta\umunu +(\partial\umu\phi^i)(\partial\unu\phi^j)\delta_{ij}=(\partial\umu x^a)(\partial\unu x^b)\eta_{ab}.
\end{align}
Note that the background metric $g\umunu(x)$ is \emph{not} the metric responsible for the gravitational coupling in the action, since there $g\umunu$ will enter only implicitly. Moreover, all fields that couple to such a background will live on the brane $\cM^{2n}$ only. In contrast to braneworld-scenarios, in this model there is no higher-dimensional ``bulk'' that could carry any physical degrees of freedom.  

\paragraph{Matrix model coordinates.} 
Throughout this work we work with so called ``matrix coordinates''
which are preferred coordinates $x^\mu \sim X^\mu$
in the model. They are such that  in the case $D=4$
the background metric is 
given by $g\umunu = \eta\umunu$ resp. $g\umunu = \delta\umunu$. 
In the general case of extra dimensions the model allows a 
$SO(D-1)$ resp. $SO(D)$ 
rotation such that at some given point $p\in\cM$ the background 
metric is again $g\umunu = \eta\umunu$ resp. $g\umunu = \delta\umunu$,
see Sect. \ref{sec: normal coordinates}.  
 
\paragraph{Geometry arises dynamically.} 
In the above matrix model Eq. (\ref{eq: the model}) a priori there is no geometry, all we have is matrices. The geometry of this model arises dynamically. The matrix model is a theory of space-time itself, in the sense that the physically realized geometry has to fulfill the equations of motion (e.o.m.) of the theory which are given by
\begin{align}
[X^a, [X^b, X^{a\oprime}]]g_{aa\oprime}=0.
\end{align}
In the semi-classical limit these are given in matrix coordinates by~\cite{Steinacker:2008ri}
\begin{align}
\theta^{\mu\alpha}(\partial\umu\theta^{\nu\beta})g\ualphabeta \label{eom2}
&=- \theta^{\mu\alpha}\theta^{\nu\beta}(\partial\umu g\ualphabeta),\\
\Delta_{\Gtilde}\phi^i (x) =0. \label{eom}
\end{align}
Here $\Gtilde$ is given by Eq. (\ref{eq: Gtilde}) and $\Delta_{\Gtilde}$ is the Laplace-Beltrami operator.
Note that at the semi-classical level we have an harmonic embedding condition for the embedding scalar fields. 

The most prominent example for such a space-time is the 4-dimensional Moyal-Weyl plane, which is a flat manifold with
\begin{eqnarray}
[X\omu,X\onu]=&\ii\bar{\theta}\omunu, \quad &\mu,\nu=0,\hdots,3 \nonumber \\
\phi^i(X) =&0, \quad & i=1,\hdots,D-4,
\end{eqnarray}
where $\bar{\theta}$ is constant. 
However, in general the solutions will fulfill 
\begin{align}
\begin{split}
[X\omu,X\onu]&=\ii\theta\omunu(X), \\
\phi^i &= \phi^i(X).
\end{split}
\end{align}
They describe a dynamical, noncommutative and non-flat 4-dimensional manifold 
with nontrivial embedding in 10 dimensions. 

\paragraph{Effective metric.}
To understand the effective geometry on $\cM$ we couple a scalar field 
as a particle on $\cM$ to the matrix model. The only reasonable kinetic term is 
\begin{align}
S[\varphi]=-(2\pi)^n\Tr [X^a,\varphi][X^b,\varphi]\eta_{ab},
\end{align} 
which becomes in the semi-classical limit~\cite{Steinacker:2007dq}
\begin{align}
S[\varphi]\sim \int\ad^{2n}x \vert \Gtilde \umunu\vert^{1/2}\Gtilde\omunu (\partial\umu\varphi)(\partial\unu\varphi).
\end{align}
Now we can see that in fact it is the effective metric
\begin{align}
\Gtilde\omunu(x)=\ems \theta^{\mu\alpha}(x)\theta^{\nu\beta}(x)g\ualphabeta(x) 
\end{align}
which is responsible for the gravitational coupling. $g\ualphabeta(x)$ is the metric (\ref{eq: background metric}) induced on 
$\cM \in \R^D$ via pull-back on $g_{ab}$ and
\begin{align}
\ems &= \rho(x) \vert g\umunu \vert^{-1/2},
\end{align}
where $\rho(x)$ is stated in Eq. (\ref{eq: Gtilde}). 
Note that the matrix model action Eq. (\ref{eq: the model}) with $\Psi=\bar{\Psi}=0$ can be written in the semi-classical limit as
\begin{align}\label{eq: bare matrix model action}
S_{YM}=-(2\pi)^n\,\Tr \frac 14 [X^a,X^b][X^{a^\prime},X^{b^\prime}]\eta_{aa^\prime}\eta_{bb^\prime}
\sim \int \ad^{2n} x \rho(x) \eta(x),
\end{align}
where 
\begin{align}
\eta(x)=\frac{1}{4} \eps \Gtilde\omunu g\umunu. 
\end{align}
\paragraph{Self-dual solutions and $\Gtilde\umunu = g\umunu$.}
Using \eq{eom}, the e.o.m. (\ref{eom2}) can be written in 
covariant form~\cite{Steinacker:2008ri},
\begin{align}\label{eq: eom covariant}
\Gtilde\omunu\widetilde{\nabla}\umu\left( \eps \thetainv_{\nu\rho} \right)
=\ems\Gtilde_{\rho\mu}\theta\omunu\partial\unu\left( \eta(x)\right).
\end{align}
A simple but important class of solutions of this equation is
given by 2-forms $\thetainv\umunu$ satisfying
\begin{align}
\label{gequalG}
\Gtilde\umunu = g\umunu. 
\end{align}
In this case, \eq{eq: eom covariant} simplifies to 
\be
g\omunu \nabla\umu \thetainv_{\nu\alpha} =0 .
\ee
It is not hard to see that in 4 dimensions, \eq{gequalG}
is equivalent to self-dual $\theta^{\mu\nu}$ \cite{Steinacker:2009mp}. 
Such solutions are of great interest in this framework because
they correlate to the cosmological constant
problem. In that case the bare matrix model action Eq.(\ref{eq: bare matrix model action}) becomes
\begin{align}
S_{YM} = \int \ad^4 x \,\rho(x)\,\eta(x) =
\int \ad^4 x \rho \eps =
 \int \ad^4 x \sqrt{\vert g\umunu \vert},
\end{align}
which is precisely the form of the induced vacuum energy interpreted as cosmological constant in GR. 
Now the variation of this term 
\begin{align}
\delta \int \ad^4 x \sqrt{\vert g \vert} \sim
\frac{1}{4}\int \ad^4 x  \sqrt{\vert g\vert} g\omunu \delta g\umunu 
\sim
\frac{1}{4}\int \ad^4 x \sqrt{\vert g\vert}\delta \phi^i  \Delta_g \phi^j \delta_{ij}
\end{align}
vanishes for harmonic embedding Eq. (\ref{eom}). Then the 
coefficient of this term is irrelevant, and 
harmonically embedded branes are protected from the
cosmological constant problem. 
Therefore the term $\int \ad^4 x \sqrt{\vert g\vert}$ should not be interpreted
as a cosmological constant, but as a brane tension.

\section{Fermions}

The obvious way to include fermions into the matrix model is 
through the action \eq{eq: the model}. This not only provides
the appropriate coupling to the metric $\tilde G^{\mu\nu}$ \cite{Klammer:2008df}, 
it is also dictated by supersymmetry \cite{Ishibashi:1996xs}. 
The quantization of such fermions  
has been studied in the 4-dimensional model
in~\cite{Klammer:2008df} from the point of view of 
noncommutative emergent gravity.
There it was shown that the  action
\begin{align}
S[\Psi]&=(2\pi)^2\,\Tr\,\bar{\Psi}\gamma_\mu[X^\mu,\Psi]
\end{align} 
indeed induces the Einstein-Hilbert action at the quantum level, along with a
dilaton-like term. The purpose of the present work is to show how the results in~\cite{Klammer:2008df} can be generalized to the case of extra dimensions.  
We split the action according to Section \ref{par: embedding}
\begin{align}\label{eq: fermion action}
\begin{split}
S[\Psi]&=(2\pi)^n\,\Tr\,\bar{\Psi}\gamma_a[X^a,\Psi] \\
&= (2\pi)^n\,\Tr\,(\bar{\Psi}\gamma_\mu[X^\mu,\Psi]+\bar{\Psi}\gamma_i[X^i,\Psi]) \\
&\sim \int \ad^{2n} x\, \rho(x) 
\bar{\Psi}\ii (\gamma\umu + \gamma_{3+i}\partial\umu \phi^i)\theta\omunu(x)\partial\unu \Psi \\
&=\int \ad^{2n}x\,\rho(x)\bar{\Psi}\ii \widetilde{\gamma}\umu\theta\omunu \partial\unu \Psi, 
\end{split}
\end{align}
where $\gamma_a$ denotes the $D$-dimensional Euclidean Clifford algebra. 
We have introduced the ``tangential'' Clifford algebra associated with the background metric $g\umunu(x)$ on $\cM$ 
whose elements are denoted by
 \begin{align}
\widetilde{\gamma}_{\,\mu}(x) = \left(\gamma\umu + \gamma_{3+i}\partial\umu \phi^i\right),
\end{align}
 and which satisfies
\begin{align}
\left\{\widetilde{\gamma}^{\,\mu},\widetilde{\gamma}^{\,\nu} \right\}
= 2 \left(\eta\urhosigma + 2 (\partial\urho\phi^i)(\partial\usigma\phi^j)\delta_{ij}\right)
=2  g\umunu(x) .
\end{align}
Notice that the $\gammat(x)$-matrices are functions of $x$, and
related to $\gamma\umu$ via some vielbein relating the 
tangent space to the ambient $\R^D$; in particular, 
$\widetilde{\gamma}_{\,\mu}(x) = \gamma\umu$ in normal coordinates 
\eq{normal-coord-1}.
The (matrix) Dirac operator is then given by
\begin{align}
\begin{split}\label{eq: Dirac op}
\slashed{D}\Psi &= \gamma_a \left[Y^a, \Psi\right] 
\sim \ii\left( \gamma\umu + \gamma_{3+i} \left(\partial\umu \phi^i\right)\right)\theta\omunu(y)\partial\unu \Psi \\
&\equiv \ii \widetilde{\gamma}\umu \theta\omunu \partial\unu \Psi.
\end{split}
\end{align}
As it has already been pointed out in~\cite{Klammer:2008df} the above
result does not quite 
match with the standard covariant derivative for spinors~\cite{weinberg}
\begin{align}
\slashed{D}_{\mathrm{comm}}\Psi=\ii \gamma^a e_a\omu \left(\partial\umu + \Sigma_{bc}\omega^{bc}\umu\right)\Psi,
\end{align}
where
\begin{align}
\omega^{ab}\umu = \frac{\ii}{2}e^{a\nu}\nabla\umu e^b\unu
\end{align}
is the usual spin connection, and
$\Sigma_{ab}=\frac{\ii}{4}[\gamma_a,\gamma_b]$. 
While the explicit derivative term is essentially the same, 
the spin connection vanishes in matrix coordinates $x^\mu$. 
This means that fermions (as long as they can be considered as 
point particles) move along geodesics of $\Gtilde\omunu$ as expected,
however with a non-standard gravitational ``spin-dragging''.
A further remark is in order. In contrast to~\cite{Klammer:2008df}, in the case of extra dimensions the Poisson structure $\theta\omunu$ does not play the sole role of a vielbein,  
 rather it is part of a  vielbein structure composed of $\theta\omunu$
 and $\partial\umu \phi^i$. 

It is easy to show that the corresponding effective metric for fermions
\begin{align}
\Gtilde\omunu_{\tau}(x)=e^{-\tau}\theta^{\mu\rho}\theta^{\nu\sigma}g\umunu(x), 
\end{align}
comes with an unusual scaling factor $e^{-\tau}$,
\begin{align}\label{eq: rescaled fermionic action}
  e^{-\tau} = \vert G\umunu \vert^{1/6} \vert g\umunu \vert^{-1/6}
= e^{-\frac{2}{3}\sigma},
\end{align} 
where 
\begin{align}
G\omunu(x)=\theta^{\mu\alpha}(x)\theta^{\nu\beta}(x)g\ualphabeta(x).
\end{align}
The scaling factor is such that it gives the correct density factor 
$\sqrt{|\Gtilde\omunu_{\tau}|}$ in the action (\ref{eq: fermion
  action}). 


\section{Quantization}
Starting from the action
\begin{align}
S[\Psi]&= (2\pi)^n\,\Tr\,\bar{\Psi}\gamma_a[X^a,\Psi]
\end{align}
we want to study the quantization of the above matrix model via a path integral over all $\Psi$, 
\begin{align}
\begin{split}
e^{-\Gamma_\Psi}&=\int \ad\Psi\ad\bar{\Psi} e^{-S[\Psi]} \\
&=\exp(\ln\det(\slashed{D}) )
= \exp \left( \frac{1}{2}\log \det (\slashed{D})^2\right) \\
&=\exp \left( \frac{1}{2}\Tr \log (\slashed{D})^2\right). 
\end{split}
\end{align}
which gives the effective action $\Gamma_\Psi$ 
\begin{align}
\Gamma_\Psi = -\frac{1}{2}\Tr \log \slashed{D}^2 .
\end{align}
Let us consider the Euclidean case for the sake of rigor. 
The square of the Dirac operator takes the following form
\begin{align}\label{eq: squared Dirac operator}
\begin{split}
\slashed{D}^2\Psi &= 
-\gammat\umu\gammat_{\rho}\theta\omunu \theta^{\rho\sigma}\partial\unu\partial_{\sigma}\Psi
-\gammat\umu\gammat_{\rho}\theta^{\mu\nu}\left(\partial\unu \theta^{\rho\sigma}\right)\left(\partial_{\sigma}\Psi\right)
-\gammat\umu \left(\partial\unu \gammat_{\rho}\right)\theta\omunu\theta^{\rho\sigma}\partial_{\sigma}\Psi \\
&=
-G\omunu\partial\umu\partial\unu \Psi - a\omu\partial\umu \Psi, 
\end{split}
\end{align}
where
$a\omu$ is the term linear in the partial derivatives
\begin{align}
\begin{split}
a^{\sigma}=\gammat\umu\gammat_{\rho}\theta\omunu\left(\partial\unu\theta^{\rho\sigma}\right)
+\gammat\umu\left(\partial\unu\gammat_{\rho}\right)\theta\omunu\theta^{\rho\sigma}.
\end{split}
\end{align}
The last term in the above equation is new in comparison
to~\cite{Klammer:2008df}, where the background $g\umunu$ was flat and
the associated Clifford algebra elements were the usual constant Dirac
matrices. To proceed, we note that
$\slashed{D}^2$ defines the  quadratic form 
\begin{align}
\begin{split}
S_{\mathrm{square}}&:=(2\pi)^n \Tr \Psi^{\dagger} \slashed{D}^2\Psi 
\sim \int\ad^{2n}x\rho(x) \Psi^{\dagger}\slashed{D}^2 \Psi \\
&=\int \ad^{2n}x \rho(x) \Psi^{\dagger}\left(-G\omunu\partial\umu\partial\unu - a\omu\partial\umu\right)\Psi\\
&=\int \ad^{2n} x \sqrt{\vert \Gtilde \vert}\Psi^{\dagger} \widetilde{\slashed{D}}^2 \Psi
\end{split}
\end{align}
which has the appropriate covariant form in terms of the metric $\Gtilde$, 
\begin{align}
\begin{split}
e^{-\sigma}&=\rho(x)\vert g\umunu(x)\vert^{-1/2}=\vert G\umunu\vert^{1/4}\vert g\umunu\vert^{-1/4} \\
\Gtilde\omunu &=e^{-\sigma} G\omunu \\
\widetilde{\slashed{D}}^2&=-\Gtilde\omunu\partial\umu\partial\unu - e^{-\sigma}a\omu \partial\unu.
\end{split}
\end{align}
In order to compute the effective action we can 
use the following integral representation of the functional determinant
\begin{align}
\begin{split}
\frac{1}{2}\Tr\left( 
\log\slashed{\widetilde{D}}^2 
\right)
&=
-\frac{1}{2}\Tr \int_0^\infty \frac{\ad\alpha}{\alpha}
\left(
e^{-\alpha\,\slashed{\widetilde{D}}^2}
\right)\\
&\equiv
-\frac{1}{2}\Tr \int_0^\infty \frac{\ad\alpha}{\alpha}
\left(
e^{-\alpha\,\slashed{\widetilde{D}}^2}
\right)e^{-\frac{1}{\alpha\Lambda^2}},
\end{split}
\end{align}
where we have introduced a cutoff $\Lambda^2$ which regularizes the divergence of $\slashed{\widetilde{D}}^2$ for small $\alpha$. 
Now we can apply the heat kernel expansion~\cite{Gilkey:1995mj,Vassilevich:2003xt} 
\begin{align}
\Tr e^{-\alpha\slashed{\widetilde{D}}^2}=\sum_{n\geq0}\alpha^{\frac{n-4}{2}}\int_{\cM}\ad^4 x\, a_n(x,\slashed{\widetilde{D}}^2)
\end{align}
where the Seeley-de Witt coefficients $a_n(y,\slashed{\widetilde{D}}^2)$ are given by
\begin{align}
\begin{split}
a_0(x)&=\frac{1}{16\,\pi^2}\tr \one,\\
a_2(x)&=\frac{1}{16\,\pi^2}\tr \left( \frac{R[\Gtilde]}{6}\one + \cE \right).
\end{split}
\end{align}
Here $\tr$ denotes the trace over the spinorial matrices and 
\begin{align}
\begin{split}
\cE&=-\Gtilde\omunu\left(
\partial\umu\Omega\unu + \Omega\umu\Omega\unu - \widetilde{\Gamma}\orho\umunu\Omega\urho
\right),\\
\Omega\umu &= \frac{1}{2}\Gtilde\umunu (\ems a\onu + \Gtilde\orhosigma\widetilde{\Gamma}\onu\urhosigma);
\end{split}
\label{e-omega}
\end{align}
note that this expression is valid only in the matrix coordinates $x^\mu$.
This gives rise to the effective action
\begin{align}
\begin{split}
\Gamma_{\Psi}=\frac{1}{16\pi^2}\int \ad^{2n}x \sqrt{\vert\Gtilde\vert}\left(
2 \tr(\one)\Lambda^4+\tr \left( \frac{R[\Gtilde]}{6}\one + \cE \right)\Lambda^2 + \cO(\log \Lambda)
\right).
\end{split}
\end{align}
This is the idea of emergent gravity observed first by Sakharov~\cite{Sakharov:1967pk}.
For the standard coupling of Dirac fermions to gravity on commutative spaces, on has~\cite{Vassilevich:2003xt}
\begin{align}
\tr \cE_{\mathrm{comm}}=-R.
\end{align}
In our case $\tr\cE$ is modified due to the non-standard spin
connection. Therefore we cannot use the standard results, and the 
geometrical meaning of \eq{e-omega} is unclear since this expression is not 
covariant and valid only in matrix coordinates.
The purpose of the present work 
is to evaluate the quantity $\tr\cE$ and see whether it gives indeed the
Ricci scalar $R[\Gtilde]$ in order to obtain the correct induced
Einstein-Hilbert 
action. We will show that $\tr\cE$ contains as expected the appropriate 
curvature scalar, plus three additional terms. This will be discussed
in detail in the following sections.

\section{Evaluation of $\tr\cE$} 

We will now determine explicitly the second Seeley-de Witt coefficient 
for the squared Dirac operator Eq. (\ref{eq: squared Dirac
  operator}). 
In order to do so we compute the Ricci scalar as well as the quantity 
$\tr\cE$ explicitly in terms of the Poisson structure, and then 
comparer those two. 
First we have to compute the following expression 
\begin{align}
\begin{split}
\tr\cE&=-\tr\left\{
\Gtilde\omunu\Omega\umu\Omega\unu + \Gtilde\omunu\partial\umu\Omega\unu -\widetilde{\Gamma}\orho\Omega\urho
\right\}\\
&=
-\tr \left(
\frac{1}{4}\Gtilde\umunu \widetilde{a}\omu \widetilde{a}\onu -\frac{1}{4}\Gtilde\umunu \widetilde{\Gamma}\omu \widetilde{\Gamma}\onu 
+\frac{1}{2}\Gtilde\omunu \partial\umu \big(
\Gtilde_{\nu\rho}\widetilde{a}\orho + \Gtilde_{\nu\rho}\widetilde{\Gamma}\urho
\big)
\right),
\end{split}
\end{align}
where $\widetilde{\Gamma}\omunu=\Gtilde\orhosigma\widetilde{\Gamma}\omu\urhosigma$.
The explicit evaluation of $\tr\cE$ is given in Appendix A. 
The result Eq. (\ref{eq: trE final}) is 
\begin{align}\label{eq: trE}
\begin{split}
\tr \cE 
&=-\tr \left(
\frac{1}{4}\Gtilde\umunu \widetilde{a}\omu \widetilde{a}\onu 
-\frac{1}{4}\Gtilde\umunu \widetilde{\Gamma}\omu \widetilde{\Gamma}\onu 
\right)\\
&=
-\ems \frac{k}{4}\Big\{
-G\omunu G\orhosigma (\partial\umu \thetainv_{\rho\alpha})(\partial\unu \thetainv_{\sigma\beta})g\oalphabeta 
+G^{\mu\nu}G^{\rho\sigma}(\partial\umu \thetainv_{\rho\alpha})(\partial\usigma \thetainv_{\nu\beta})g\oalphabeta \\
&\quad
+G^{\mu\sigma}\theta^{\rho\alpha}(\partial\usigma g_{\alpha\delta})(\partial\urho\thetainv\umunu)g^{\nu\delta}
+G^{\mu\sigma}\theta^{\rho\alpha}(\partial\udelta g_{\alpha\sigma})(\partial\urho \thetainv\umunu)g^{\nu\delta}\\
&\quad
-G^{\mu\sigma}\theta^{\rho\alpha}(\partial\ualpha g_{\sigma\delta})(\partial\urho \thetainv\umunu)g^{\nu\delta}
-G\orhosigma\theta^{\mu\beta}(\partial\usigma g_{\beta\delta})(\partial\urho \thetainv\umunu)g^{\nu\delta}\\
&\quad
-G\orhosigma\theta^{\mu\beta}(\partial\udelta g_{\beta\sigma})(\partial\urho \thetainv\umunu)g^{\nu\delta}
+G\orhosigma\theta^{\mu\beta}(\partial\ubeta g_{\sigma\delta})(\partial\urho \thetainv\umunu)g^{\nu\delta} \\
&\quad
-\frac{1}{2}\big(
G\omunu(g\partial\umu\partial\unu g\hme) 
-2G\orhosigma g^{\delta\beta}\partial\urho\partial\ubeta g_{\sigma\delta}
+g\omunu G\orhosigma \partial\umu\partial\unu g_{\rho\sigma}
\big)\\
&\quad
+\frac{1}{2}\theta^{\rho\alpha}(\partial\urho g_{\gamma\beta})\theta^{\sigma\gamma}(\partial\usigma g_{\alpha\delta})g^{\delta\beta} 
+\frac{1}{2}\theta\omunu\theta\orhosigma(\partial\umu g_{\rho\alpha})(\partial\unu g_{\sigma\beta})g\oalphabeta \\
&\quad
-\frac{1}{4}\theta\omunu\theta\orhosigma(\partial\ualpha g_{\mu\rho})(\partial\ubeta g_{\nu\sigma})g\oalphabeta
\Big\}\\
&\quad
+\frac{k}{4}g\omunu(\partial\umu\partial\unu\phi^i)
\big(
\Delta_{\Gtilde}\phi^j+\widetilde{\Gamma}\orho \partial\urho \phi^j
\big)\delta_{ij} 
+\frac{k}{4}\Gtilde\umunu \widetilde{\Gamma}\omu \widetilde{\Gamma}\onu,
\end{split}
\end{align}
where
\begin{align}
k=\mathrm{rank}(\gamma).
\end{align}
$k$ is the rank of the representation of $D$-dimensional Clifford algebra, depending on the number of extra dimensions. 

For the sake of simplicity and manageability we will sometimes use the equations of motions, Eq. (\ref{eom2}) and (\ref{eom})  and work with on-shell geometries. 
Then the contracted Christoffel symbols vanish~\cite{Steinacker:2008ri}, 
\begin{align}\label{eq: Gamma tilde null}
\begin{split}
\widetilde{\Gamma}\omu&=-\partial\urho \Gtilde^{\rho\mu}-\frac{1}{2}\Gtilde\omunu(\Gtilde\partial\unu \Gtilde^{-1}) \\
&=e^{-\sigma}\left(-(\partial\urho\theta^{\mu\beta})\theta^{\rho\alpha}g\ualphabeta - \theta^{\mu\beta}\theta^{\rho\alpha}(\partial\urho g\ualphabeta)
\right) \\
&\stackrel{\mathrm{e.o.m.}}{=}0.
\end{split}
\end{align}
Due to Eq.(\ref{eq: Gamma tilde null}) also the harmonic embedding condition simplifies as 
\begin{align}\label{eq: harmonic embedding }
\begin{split}
\Delta_{\Gtilde} \phi &= 
\left(
\Gtilde\omunu\partial\umu \partial\unu - \widetilde{\Gamma}\omu\partial\umu
\right)\phi \\
&=\Gtilde\omunu \partial\umu\partial\unu \phi \\
&=0.
\end{split}
\end{align}
These handy features simplify our calculations a lot since $\tr \cE$ is then determined by a single term,  
\begin{align}
\begin{split}
\tr\cE&=-\frac{\ems}{4} \tr\left(G\umunu a\omu a\onu\right).
\end{split}
\end{align}
In principle one could now go on and compute the Ricci scalar in terms
of the Poisson tensor $\theta\omunu$ and compare the two
quantities. This strategy was pursued
in~\cite{Klammer:2008df}. However, it turns out that this procedure is
too complicated and seems to be not feasible in the case of extra
dimensions. Hence we simplify our computations by going
to \emph{normal coordinates}, i.e. coordinates where first order
derivatives in the embedding scalar fields, $\partial\umu\phi(x)$,
vanish. But to do that, we should first show that $\tr \cE$ 
is a covariant expression. This is done in Appendix B, and we only quote the
result here.  For on-shell geometries which satisfy \eq{eom2}, \eq{eom},
we find
\begin{align}\label{eq: covariant E}
\begin{split}
\tr \cE 
 &= -\frac{\ems}{4}\,k\,
\Big(
G\omunu(\nabla\umu\thetainv_{\nu\alpha})G\orhosigma(\nabla\urho\thetainv_{\sigma\beta})g\oalphabeta 
-G\omunu G\orhosigma(\nabla\umu\thetainv_{\rho\alpha})(\nabla\unu\thetainv_{\sigma\beta})g\oalphabeta \\
&\quad 
+G\omunu G\orhosigma (\nabla\umu\thetainv_{\rho\alpha})(\nabla\usigma\thetainv_{\nu\beta})g\oalphabeta
\Big)
-\frac{k}{4}\Gtilde\omunu R\umunu[g],
\end{split}
\end{align}
see also Eq. (\ref{eq: covariant E II}) of Appendix B. 
In the special case of identical background and effective metric $\Gtilde\umunu=g\umunu$ the use of on-shell geometries is not necessary. Then we have
\begin{align}
\begin{split}
\tr \cE &=
- \frac{\ems}{4}
\tr\, G\umunu a\omu a\onu +  \frac{\ems}{4}
\tr\, g\umunu \Gamma\omu \Gamma\onu\\
 &= -\frac{\eps}{4}\,k\,
g\omunu g\orhosigma (\nabla\umu\thetainv_{\rho\alpha})(\nabla\usigma\thetainv_{\nu\beta})g\oalphabeta
-\frac{k}{4} R[g]
+\frac{k}{4}  (\Delta_g x^a)(\Delta_g x^b)\eta_{ab},
\end{split}
\label{eq: covariant E-2}
\end{align}
as stated in Eq. (\ref{eq: covariant E III}). This expression has a
clear geometrical meaning (taking into account extrinsic geometry 
in the last term) and can thus be considered as covariant.
For on-shell geometries, the last term vanishes and \eq{eq: covariant E-2}
agrees with \eq{eq: covariant E} using the Bianci identity \eq{bianci}.

\paragraph{A short remark regarding notation.} We have to distinguish between the \emph{effective metric} $\Gtilde\umunu$ and the \emph{background metric} $g\umunu$. Covariant derivatives and Christoffel symbols with respect to the effective metric $\Gtilde\umunu$ in NC emergent gravity are usually denoted as 
$\widetilde{\nabla}\umu$ and $\widetilde{\Gamma}\omu\urhosigma$, respectively.  Covariant derivatives and Christoffel symbols with respect to the background metric $g\umunu$ as they appear in Eq.(\ref{eq: covariant E}) are written as $\nabla\umu$ and $\Gamma\omu\urhosigma$.


\section{Going to a normal embedding coordinate system}\label{sec: normal coordinates}
\subsection{$\tr\cE$ in normal coordinates.}
Since the matrix model action is invariant under $SO(D)$
resp. $SO(1,D-1)$ rotations as well as translations,
one can choose for any given point $p \in \cM$ adapted coordinates such that the brane is tangential to the plane spanned by the first $2n$ components. Then we have at this point 
\begin{align}
\partial\umu \phi^i \vert_p &=0, \label{normal-coord-1}\\
\partial\umu g\urhosigma \vert_p&=0.  
\end{align}
We denote such coordinates as ``normal embedding coordinates'' or simply ``normal coordinates''. They are still matrix coordinates $x^a \sim X^a$ and thus the e.o.m. Eq. (\ref{eom2}) and (\ref{eom}) still hold.   
We can now take our result of Eq. (\ref{eq: trE}) for $\tr\cE$ and
write it in normal coordinates by simply omitting all terms with first
partial derivatives of the background metric $g\umunu$. 
Since the covariance of $\tr\cE$ of Eq. (\ref{eq: covariant E}) for general $\Gtilde$ is
only established by making use of the e.o.m., we have to work with on-shell geometries. $\tr\cE$ in normal coordinates is thus given by
\begin{align}
\begin{split}
\tr\cE 
&=-\frac{\ems}{4}\Tr\left\{G\umunu a\omu a\onu \right\}\\
&=\ems k\,\Big\{
\frac{1}{4}G\omunu G\orhosigma(\partial\umu\thetainv_{\rho\alpha})(\partial\unu\thetainv_{\sigma\beta})\eta\oalphabeta
-\frac{1}{4}G\omunu G\orhosigma(\partial\umu\thetainv_{\rho\alpha})(\partial\usigma\thetainv_{\nu\beta})\eta\oalphabeta \\
&\quad
+\frac{1}{8}\big(
G\omunu(g\orhosigma\partial\umu\partial\unu g\urhosigma)-2G\orhosigma g^{\delta\beta}\partial\urho\partial\ubeta g_{\sigma\delta}
+g\omunu G\orhosigma \partial\umu\partial\unu g_{\rho\sigma} 
\big)\Big\} \\
&\quad
+\frac{k}{4}g\omunu(\partial\umu\partial\unu\phi^i)\left(
\Delta_{\Gtilde}\phi^j+\widetilde{\Gamma}\orho(\partial\urho\phi^j)\delta_{ij}
\right)
+\frac{k}{4}\Gtilde\umunu \widetilde{\Gamma}\omu \widetilde{\Gamma}\onu
.
\end{split}
\end{align}
We make use of the relation Eq.(\ref{relation phi Gamma}) of Appendix A 
\begin{align}
\begin{split}
(\partial\ulambda\phi^i)(\partial\umu\partial\unu\phi^j)\delta_{ij}
&=
\frac{1}{2}\big(
\partial\umu g_{\nu\lambda}+\partial\unu g_{\mu\lambda}-\partial\ulambda g\umunu
\big).
\end{split}
\end{align}
Differentiating once more gives
\begin{align}
\begin{split}\label{eq: ur-wichtige relation}
(\partial\urho\partial\usigma\phi^i)(\partial\umu\partial\unu\phi^j)\delta_{ij}
&\stackrel{\mathrm{nc}}{=}\frac{1}{2}\left(
\partial\urho\partial\umu g_{\nu\sigma}+\partial\urho\partial\unu g_{\mu\sigma}-\partial\urho\partial\usigma g\umunu
\right)
\end{split}
\end{align}
where the superscript ``nc'' stands for normal coordinates.
In normal coordinates we have
\begin{align}\label{eq: a metric relation}
\begin{split}
g\orhosigma G\omunu (\partial\urho\partial\umu g_{\nu\sigma})&=
g\orhosigma(\partial\urho\partial\usigma\phi^i)G\omunu(\partial\umu\partial\unu\phi^j)\delta_{ij}
+\frac{1}{2}g\omunu G\orhosigma (\partial\urho\partial\usigma g\umunu)\\
&=
\eps g\orhosigma(\partial\urho\partial\usigma\phi^i)
\left( \Delta_{\Gtilde}\phi^j +\widetilde{\Gamma}\omu(\partial\umu\phi^j)\right)\delta_{ij}
+\frac{1}{2}g\omunu G\orhosigma (\partial\urho\partial\usigma g\umunu).
\end{split}
\end{align}
Our final result for $\tr\cE$ in normal coordinates is then
\begin{align}
\begin{split}
\tr\cE 
&=\ems \frac{k}{4}\,\Big\{
G\omunu G\orhosigma(\partial\umu\thetainv_{\rho\alpha})(\partial\unu\thetainv_{\sigma\beta})g\oalphabeta
-G\omunu G\orhosigma(\partial\umu\thetainv_{\rho\alpha})(\partial\usigma\thetainv_{\nu\beta})g\oalphabeta \\
&\quad
+\frac{1}{8}G\omunu(g\orhosigma\partial\umu\partial\unu g\urhosigma) 
\Big\} 
+\frac{k}{4}\Gtilde\umunu \widetilde{\Gamma}\omu \widetilde{\Gamma}\onu \\
&\stackrel{\mathrm{e.o.m.}}{=}
\ems \frac{k}{4}\,\Big\{
G\omunu G\orhosigma(\partial\umu\thetainv_{\rho\alpha})(\partial\unu\thetainv_{\sigma\beta})g\oalphabeta
-G\omunu G\orhosigma(\partial\umu\thetainv_{\rho\alpha})(\partial\usigma\thetainv_{\nu\beta})g\oalphabeta \\
&\quad
+\frac{1}{8}G\omunu(g\orhosigma\partial\umu\partial\unu g\urhosigma) 
\Big\}.
\end{split}
\end{align}

\subsection{$\tr\cE$ in normal coordinates for $\Gtilde = g$}

Since it was shown in the last section that $\tr\cE$ for $\Gtilde = g$ is a covariant expression even for off-shell geometries, we will not use the e.o.m. here. 
The term $\frac{k}{4}g\umunu \Gamma\omu \Gamma\onu$ now vanishes due to the normal coordinate system.
Since $\thetainv\umunu$ fulfills the Jacobi identity the following equation 
\begin{align}
2(\partial\umu\thetainv_{\rho\alpha})(\partial\usigma\thetainv_{\nu\beta})g\omunu g\orhosigma g\oalphabeta 
=(\partial\umu\thetainv_{\rho\alpha})(\partial\unu\thetainv_{\sigma\beta})g\omunu g\orhosigma g\oalphabeta 
\end{align}
holds for $\Gtilde=g$, see also Appendix B. This simplifies $\tr\cE$ to
\begin{align}
\begin{split}
\tr \cE &= 
 \eps\frac{k}{4}
g\omunu g\orhosigma (\partial\umu \thetainv_{\rho\alpha})(\partial\usigma \thetainv_{\nu\beta})g\oalphabeta
+\frac{k}{8}g\omunu (g\orhosigma\partial\umu\partial\unu g\urhosigma).
\end{split}
\end{align}
\subsection{The Ricci scalar $R[\Gtilde]$ in normal coordinates.}
Let us now study the Ricci scalar $R[\Gtilde]$ in normal coordinates. The curvature tensor 
and the Ricci scalar are given as usual by
\begin{align}
\begin{split}
R_{\mu\nu\rho}^{\phantom{\mu\nu\rho}\sigma}[\Gtilde]&=
\partial\unu \widetilde{\Gamma}\osigma_{\mu\rho}
-\partial\umu \widetilde{\Gamma}_{\nu\rho}\osigma
+\widetilde{\Gamma}\olambda_{\mu\rho}\widetilde{\Gamma}\osigma_{\lambda\nu}
-\widetilde{\Gamma}\olambda_{\nu\rho}\widetilde{\Gamma}\osigma_{\lambda\mu},\\
R[\Gtilde]&=\Gtilde^{\mu\rho}R_{\mu\nu\rho}^{\phantom{\mu\sigma\rho}\nu}.
\end{split}
\end{align}
In terms of the metric (now with respect to the effective metric $\Gtilde$) and its derivatives the Ricci scalar is given by
\begin{align}\label{eq: R}
\begin{split}
R[\Gtilde]&=-\Gtilde\umunu(\partial\urho \Gtilde^{\rho\mu})(\partial\usigma\Gtilde^{\sigma\nu})
+\Gtilde\omunu \Gtilde\orhosigma (\partial\umu\partial\urho \Gtilde_{\nu\sigma}) \\
&\quad
-\Gtilde\omunu\Gtilde\orhosigma\partial\urho\partial\usigma \Gtilde\umunu 
-(\partial\urho \Gtilde\orhosigma)(\Gtilde\omunu\partial\usigma \Gtilde\umunu) \\
&\quad
-\frac{3}{4}\Gtilde\omunu(\partial\umu \Gtilde\orhosigma)(\partial\unu \Gtilde\urhosigma)
+\frac{1}{2}\Gtilde\orhosigma(\partial\usigma\Gtilde\omunu)(\partial\unu \Gtilde_{\mu\rho}) \\
&\quad
-\frac{1}{4}\Gtilde\omunu(\Gtilde\orhosigma\partial\umu \Gtilde\urhosigma)(\Gtilde^{\kappa\lambda}\partial\unu\Gtilde_{\kappa\lambda}). 
\end{split}
\end{align}
\paragraph{$R[\Gtilde]$ for on-shell geometries $\Gtilde \neq g$ using e.o.m.}
See Appendix C for the evaluation of the Ricci scalar in normal coordinates.
The result is found to be
\begin{align}
\begin{split}
R[\Gtilde]&\stackrel{nc}{=}
\ems\Big\{
\frac{1}{2}(\partial\umu\theta^{\mu\alpha})(\partial\unu\theta^{\nu\beta})\eta\ualphabeta
+\frac{1}{2}(\partial\umu\theta^{\nu\alpha})(\partial\unu\theta^{\mu\beta})\eta\ualphabeta\\
&\quad
+\frac{1}{2}G\omunu G\orhosigma(\partial\umu\thetainv_{\rho\alpha})(\partial\usigma\thetainv_{\nu\beta})\eta\oalphabeta
-\frac{1}{2}G\omunu G\orhosigma(\partial\umu\thetainv_{\rho\alpha})(\partial\unu\thetainv_{\sigma\beta})\eta\oalphabeta\\
&\quad
-\frac{1}{2}G\omunu(g\orhosigma\partial\umu\partial\unu g\urhosigma)
\Big\}.
\end{split}
\end{align}
\paragraph{$R[\Gtilde]$ for self-dual off-shell geometries $\Gtilde = g$.}
If the background metric equals the effective metric, $\Gtilde\umunu=g\umunu$, the Ricci scalar in normal coordinates is due to Eq.(\ref{eq: a metric relation})
\begin{align}
\begin{split}
R[g]&=g\omunu g\orhosigma (\partial\umu\partial\urho g_{\nu\sigma})
-g\omunu g\orhosigma\partial\urho\partial\usigma g\umunu \\
&=-\frac{1}{2}g\omunu g\orhosigma\partial\urho\partial\usigma g\umunu
+(\Delta_g \phi^i)(\Delta_g \phi^j)\delta_{ij}.
\end{split}
\end{align}
Hence, in that special case we have
\begin{align}
\begin{split}
(\partial\umu\theta^{\mu\alpha})(\partial\unu\theta^{\nu\beta})g\ualphabeta
+(\partial\umu\theta^{\nu\alpha})(\partial\unu\theta^{\mu\beta})g\ualphabeta
&=
-e^{2\sigma}g\omunu g\orhosigma(\partial\umu\thetainv_{\rho\alpha})(\partial\usigma\thetainv_{\nu\beta})g\oalphabeta \\
&\quad
+e^{2\sigma}g\omunu g\orhosigma(\partial\umu\thetainv_{\rho\alpha})(\partial\unu\thetainv_{\sigma\beta})g\oalphabeta \\
&=
e^{2\sigma}g\omunu g\orhosigma(\partial\umu\thetainv_{\rho\alpha})(\partial\usigma\thetainv_{\nu\beta})g\oalphabeta.
\end{split}
\end{align}
\subsection{A comparison of $\tr\cE$ \& $R[g]$}\label{sect: comparison}
Let us finally compare our results for $\tr\cE$ and the Ricci scalar $R[\Gtilde]$.
In normal coordinates we find the following relation between the Ricci scalar and $\tr\cE$. 
\begin{align}
\begin{split}
\tr\cE &= -\frac{k}{2}R[\Gtilde] - \frac{k}{8}G\omunu(g\orhosigma\partial\umu\partial\unu g\urhosigma) \\
&\quad
+\frac{k}{4}\ems (\partial\umu\theta^{\mu\alpha})(\partial\unu\theta^{\nu\beta})g\ualphabeta
+\frac{k}{4}\ems (\partial\umu\theta^{\nu\alpha})(\partial\unu\theta^{\mu\beta})g\ualphabeta 
\end{split}
\end{align}
Let us write this result again as a covariant expression. In order to do so, we notice that 
\begin{align}
\begin{split}
\theta^{\mu\alpha}(\nabla\umu\theta^{\nu\beta})g\ualphabeta &=
\theta^{\mu\alpha}(\partial\umu\theta^{\nu\beta})g\ualpha + \theta^{\mu\alpha}\Gamma\onu_{\mu\rho}\theta^{\rho\beta}g\ualphabeta
+\theta^{\mu\alpha}\Gamma\obeta_{\mu\rho}\theta^{\nu\rho}g\ualphabeta \\
&=
\theta^{\mu\alpha}(\partial\umu\theta^{\nu\beta})g\ualpha + \Gamma\onu\urhosigma G\orhosigma
+\theta^{\mu\alpha}\theta^{\nu\rho}(\partial\umu g_{\alpha\rho})\\
&=
\Gamma\onu\urhosigma G\orhosigma \\
&=G\orhosigma (\partial\urho g_{\lambda\sigma})g^{\lambda\nu}
-\frac{1}{2}G\orhosigma g^{\lambda\nu}(\partial\ulambda g\urhosigma)\\
&=0,
\end{split}
\end{align}
using the e.o.m and Eq. (\ref{eq: ur-wichtige relation}). A consequence of the above relation is then
\begin{align}
\begin{split}
(\nabla\umu \theta^{\nu\alpha})(\nabla\unu \theta^{\mu\beta})g\ualphabeta =
-\theta^{\mu\alpha}(\nabla\unu\nabla\umu \theta^{\nu\beta})g\ualphabeta.
\end{split}
\end{align}
In normal coordinates this is 
\begin{align}
\begin{split}
(\partial\umu\theta^{\nu\alpha})(\partial\unu\theta^{\mu\beta})g\ualphabeta
=-\theta^{\mu\alpha}(\partial\umu\partial\unu \theta^{\nu\beta})g\ualphabeta
-\theta^{\mu\alpha}\theta^{\nu\beta}\partial\umu\partial\unu g\ualphabeta,
\end{split}
\end{align}
which can also be derived from the equation of motion. 
Now remember that
\begin{align}
\begin{split}
\theta^{\mu\alpha}(\nabla\unu\nabla\umu\theta^{\nu\beta})g\ualphabeta
&=
-\theta^{\mu\alpha}(\nabla\umu\nabla\unu\theta^{\nu\beta})g\ualphabeta
+R\omu_{\lambda\mu\nu}G^{\lambda\nu}
+R\obeta_{\lambda\mu\nu}\theta^{\mu\lambda}\theta^{\nu\alpha}g\ualphabeta \\
&=
-\theta^{\mu\alpha}(\nabla\umu\nabla\unu\theta^{\nu\beta})g\ualphabeta
+G\omunu R[g]\umunu
-\frac{1}{2} R[g]_{\mu\nu\rho\sigma}\theta\omunu \theta\orhosigma.
\end{split}
\end{align}
Next consider the Ricci tensor in normal coordinates. 
\begin{align}
\begin{split}
R\umunu
&\stackrel{nc}{=}
\partial\urho\Gamma\orho_{\mu\nu}-\partial\umu\Gamma\orho_{\rho\nu}\\
&=
\frac{1}{2}g^{\rho\lambda}\Big(
2\partial\umu\partial\urho g_{\lambda\nu}
+\partial\unu\partial\urho g_{\mu\lambda}-\partial\urho \partial\ulambda g_{\mu\nu}
+\partial\umu\partial\unu g_{\rho\lambda}-\partial\umu\partial\ulambda g_{\nu\rho}
\Big)
\end{split}
\end{align}
Due to Eq.(\ref{eq: a metric relation}) we find then
\begin{align}
\begin{split}
G\omunu R[g]\umunu \stackrel{nc}{=} -\frac{1}{2}G\omunu (g\orhosigma\partial\umu\partial\unu g\urhosigma).
\end{split}
\end{align}
Using also
\begin{align}
\begin{split}
\theta^{\mu\alpha}(\nabla\umu\nabla\unu\theta^{\nu\beta})g\ualphabeta
=G\omunu \partial\umu\partial\unu\sigma
=G\omunu\nabla\umu\nabla\unu \sigma
\end{split}
\end{align}
as well as
\begin{align}
\begin{split}
(\nabla\umu \theta^{\mu\alpha})(\nabla\unu\theta^{\nu\beta})g\ualphabeta \stackrel{nc}{=}
(\partial\umu \theta^{\mu\alpha})(\partial\unu\theta^{\nu\beta})g\ualphabeta
=G\omunu(\partial\umu\sigma)(\partial\unu\sigma)
=G\omunu(\nabla\umu\sigma)(\nabla\unu\sigma)
\end{split}
\end{align}
we obtain the following covariant form of $\tr\cE$,
\begin{align}
\tr\cE =
-\frac{k}{2}R[\Gtilde]
+\frac{k}{4}\Gtilde\omunu(\nabla\umu\sigma)(\nabla\unu \sigma)
+\frac{k}{4}\Gtilde\omunu\nabla\umu\nabla\unu \sigma
+\frac{k}{8}\,\ems\,R[g]_{\mu\nu\rho\sigma}\theta\omunu \theta\orhosigma
\end{align}
or
\begin{align}\label{eq: result I}
\tr\cE =
-\frac{k}{2}R[\Gtilde]
+\frac{k}{4}\ems \Delta_{\tilde G} \eps
+\frac{k}{8}\,\ems\,R[g]_{\mu\nu\rho\sigma}\theta\omunu \theta\orhosigma.
\end{align}
As a check, we can compare this with the result
of~\cite{Klammer:2008df}, where
the case of a 4-dimensional manifold with flat background metric was
studied. $\tr\cE$ for on-shell geometries was shown to be
\begin{align}
\int \ad^4 x \; \tr\cE = \int \ad^4 x \,k_{4d}\,\left( -\frac{1}{2}R[\Gtilde] + G\omunu (\partial\umu\sigma)(\partial\unu\sigma)\right),
\end{align}
where the rank $k_{4d}$ of the 4-dimensional representation of the Clifforda algebra is four.
Hence the two results are consistent if the background
metric is flat and the number of dimensions equals four. 

Finally for on-shell geometries we find the following one-loop effective action
\begin{align}
\Gamma_{\Psi} &= \frac{k}{16 \pi^2}\int \ad^{2n}x \sqrt{\vert\Gtilde\vert}
\Big(
2  \Lambda^4 +
\big(
-\frac{1}{3}R[\Gtilde] + \frac{1}{4}\ems\Delta_{\tilde G} \eps \nn\\
&\quad
+\frac{1}{8}\ems R[g]_{\mu\nu\rho\sigma}\theta\omunu \theta\orhosigma
\big)\Lambda^2
+\cO(\log \Lambda)
\Big).
\label{eff-action-final}
\end{align}

\paragraph{A comparison of $\tr\cE$ and $R$ for $\Gtilde = g$ without using the e.o.m.}
\begin{align}
\tr\cE&\stackrel{nc}{=}\frac{k}{4}\ems\left(
(\partial\umu\theta^{\mu\alpha})(\partial\unu\theta^{\nu\beta})g\ualphabeta
+(\partial\umu\theta^{\nu\alpha})(\partial\unu\theta^{\mu\beta})g\ualphabeta
\right)
-\frac{k}{4}R[g]\\
&\quad
+\frac{k}{4}(\Delta_g\phi^i)(\Delta_g\phi^j)\delta_{ij}.
\end{align}
Recall that
\begin{align}
\begin{split}
\theta^{\mu\alpha}(\nabla\umu\theta^{\nu\beta})g\ualphabeta 
&=
\theta^{\mu\alpha}(\partial\umu\theta^{\nu\beta})g\ualphabeta + G^{\mu\rho}\Gamma\onu_{\mu\rho}
+\theta^{\mu\alpha}\theta^{\nu\beta}(\partial\unu g\ualphabeta)\\
&\stackrel{\Gtilde = g}{=}-\eps \Gamma\onu + \eps \Gamma\onu \\
&=0
\end{split}
\end{align}
So 
\begin{align}
(\nabla\umu\theta^{\nu\alpha})(\nabla\unu\theta^{\mu\beta})g\ualphabeta =
-\theta^{\mu\alpha}(\nabla\unu\nabla\umu\theta^{\nu\beta})g\ualphabeta
\end{align}
 is true also for off-shell geometries in the case of $\Gtilde = g$. 
With the help of Eq. (\ref{eq: a metric relation}) we see that
\begin{align}
\begin{split}
\theta^{\mu\alpha}(\nabla\unu\nabla\umu\theta^{\nu\beta})g\ualphabeta &\stackrel{nc}{=}
\theta^{\mu\alpha}(\partial\umu\partial\unu\theta^{\nu\beta})g\ualphabeta
+\theta^{\mu\alpha}\theta^{\nu\beta}\partial\umu\partial\unu g\ualphabeta
+(\Delta_g\phi^i)(\Delta_g\phi^j)\delta_{ij}\\
&\stackrel{nc}{=}
-(\partial\umu\theta^{\nu\alpha})(\partial\unu\theta^{\mu\beta})g\ualphabeta.
\end{split}
\end{align}
Using Eq. (\ref{eq: homogeneous maxwell}) of Appendix B we have
\begin{align}
\begin{split}
\theta^{\mu\alpha}(\nabla\umu\nabla\unu\theta^{\nu\beta})g\ualphabeta &=
\theta^{\mu\alpha}(\partial\umu\partial\unu\theta^{\nu\beta})g\ualphabeta
+\frac{\eps}{2}g\omunu g\orhosigma \partial\umu\partial\unu g\urhosigma
\stackrel{nc}{=}\eps g\omunu \partial\umu\partial\unu \sigma.
\end{split}
\end{align}

\begin{align}
\begin{split}
\tr\cE = 
-\frac{k}{2}R[g] +\frac{k}{4}\ems g\omunu \nabla\umu \nabla\unu \eps +
\frac{k}{8}\ems R[g]_{\mu\nu\rho\sigma}\theta\omunu \theta\orhosigma
+\frac{k}{4}(\Delta_g x^a)(\Delta_g x^b)\eta_{ab}.
\end{split}
\end{align}
The total one-loop effective action is thus 
\begin{align}
\Gamma_{\Psi} &= \frac{k}{16 \pi^2}\int \ad^{2n}x \sqrt{\vert g \vert}
\Big(
2  \Lambda^4 +
\big(
-\frac{1}{3}R[g] + \frac{1}{4}\ems \Delta_g \eps \nn\\
&\quad
+\frac{1}{8}\ems R[g]_{\mu\nu\rho\sigma}\theta\omunu \theta\orhosigma
+\frac{1}{4}(\Delta_g x^a)(\Delta_g x^b)\eta_{ab}
\big)\Lambda^2
+\cO(\log \Lambda)
\Big).
\label{eff-action-final-2}
\end{align}
We find that for this off-shell but self-dual case the result agrees with Eq. (\ref{eff-action-final}) plus a term depending on the extrinsic geometry.

\section{Conclusion}
In this work, fermions are studied in the framework of emergent 
noncommutative gravity for the general case of branes embedded
in higher dimensions.
The model is realized via a matrix model of Yang-Mills type. 
The fermionic term in the matrix model action leads to a specific 
coupling to geometry determined by a nontrivial effective metric 
$\Gtilde\umunu$. It goes along with a vanishing spin connection in the
preferred coordinates associated with the matrix model. 
In the case of extra dimensions the vielbein is not given entirely by 
the Poisson tensor $\theta\omunu$ as in \cite{Klammer:2008df}, rather the Poisson
tensor corresponds to a vielbein which relates the effective metric to
the ``tangential'' embedding metric which in turn is non-trivial.
This is responsible for the difference with the standard case.

The resulting action in this framework shows some deviations from the 
standard fermionic coupling. We find an induced gravitational action
which includes the expected Einstein-Hilbert term with a modified
coefficient, 
as well as three additional terms. One of these additional 
terms contains explicitly the 
Poisson structure $\theta\omunu$ coupled to the Riemann tensor,
$e^{-\sigma} R_{\mu\nu\rho\sigma}\theta^{\mu\nu}\theta^{\rho\sigma}$.
This term leads to some concern because $\theta\omunu$ breaks 
Lorentz invariance. That is irrelevant in the remaining terms of the
effective action where $\theta\omunu$ enters only implicitly through 
the effective metric, however 
$e^{-\sigma}R_{\mu\nu\rho\sigma}\theta^{\mu\nu}\theta^{\rho\sigma}$
corresponds to a direct coupling of $\theta\omunu$ on the geometry, 
which breaks the local Lorentz invariance. 
Moreover, it is not small due to the $e^{-\sigma}$ factor.
While this term vanishes in flat geometries, it 
will have some impact on gravity. 
The physical significance of this effect remains to be studied.

The importance of the additional $R \theta\theta$ term
depends on the precise mechanism of gravity in the matrix model.
As shown in \cite{Steinacker:2009mp}, there seem to be 2 possibilities:
First, gravity is indeed governed by the induced gravitational 
terms in the spirit of induced gravity. 
Then the $R \theta\theta$
term may have important physical implications. 
Second, gravity is dominated by the deformed harmonic embeddings
resp. brane tension, and the cutoff $\L$ (which is 
given by the scale of $N=4$ SUSY breaking in the IKKT model)
in front of the induced gravitational terms
is much smaller than the Planck scale; this scenario is indeed
preferred and viable as shown in \cite{Steinacker:2009mp}. In that case, the novel terms
$R \theta\theta$ obtained here would have a very small impact on
gravity and lead only to minor corrections. The latter scenario
is more attractive for a variety of reasons including the 
cosmological constant problem.

In either case, we found an interesting new term in the one-loop
effective action for fermions in the matrix model, as well as the 
usual Einstein-Hilbert term with a non-standard coefficient.
This is a manifestation of the non-standard spin connection 
in the matrix model.
Clearly more work is required before the physical implications 
of this terms are fully understood.

\paragraph{Acknowledgements} We would like to express our gratitude to H. Grosse for many discussions. The work of D.K. was supported by the FWF project P20017, and the work of H.S. was supported  in part by FWF project P20017 and in part by the FWF project P21610. 


\section{Appendix A: Evaluation of $\tr \cE$}
We want to express $\tr \cE$,
\begin{align}
\begin{split}
\tr\cE&=-\tr\left\{
\Gtilde\omunu\Omega\umu\Omega\unu + \Gtilde\omunu\partial\umu\Omega\unu -\widetilde{\Gamma}\orho\Omega\urho
\right\}\\
&=
-\tr \Big(
\frac{1}{4}\Gtilde\umunu a\omu a\onu -\frac{1}{4}\Gtilde\umunu \widetilde{\Gamma}\omu \widetilde{\Gamma}\onu
+\frac{1}{2}\Gtilde\omunu \partial\umu \big(
\Gtilde_{\nu\rho}\widetilde{a}\orho + \Gtilde_{\nu\rho}\widetilde{\Gamma}^\rho
\big)
\Big).
\end{split}
\end{align}
explicitly by the Poisson tensor $\theta\omunu$ and the background metric $g\umunu$. 

To begin with we show that the following relation containing second order partial derivatives is true.
\begin{align}
\begin{split}
g^{\lambda\nu}G^{\rho\mu}
(\partial\urho\partial_{\lambda}\phi^i)(\partial\umu\partial\unu\phi^j)\delta_{ij} 
&= 
\frac{1}{2}\Big(
g\orhosigma G\omunu \partial\umu\partial\unu g\urhosigma
+ G^{\rho\lambda}g\omunu \partial\umu\partial\unu g_{\rho\lambda}
 - 2 G^{\rho\mu}g^{\lambda\nu}\partial\urho\partial_{\lambda}g\umunu
\Big)\\
&\quad
+\eps g\omunu (\partial\umu\partial\unu \phi^i)
\left(
\Delta_{\Gtilde}\phi^j+\widetilde{\Gamma}\omu (\partial\umu \phi^j)
\right)\delta_{ij}
.
\end{split}
\end{align}
This can be seen by taking 
\begin{align}
\begin{split}
(\partial\urho\partial\ubeta\phi^i)(\partial\usigma\partial\udelta\phi^j)\delta_{ij}
+(\partial\ubeta\phi^i)(\partial\urho\partial\usigma\partial\udelta\phi^j)\delta_{ij}&=
\frac{1}{2}\big(
\partial\urho\partial\usigma g_{\beta\delta} 
+\partial\urho\partial\udelta g_{\beta\sigma}\\
&\quad
-\partial\urho\partial\ubeta g_{\sigma\delta}
\big)
\end{split}
\end{align}
and subtracting from this equation the same equation with the indices $\rho$ and $\delta$ interchanged. This gives
\begin{align}
\begin{split}
(\partial\urho\partial\ubeta\phi^i)(\partial\usigma\partial\udelta\phi^j)\delta_{ij}
-(\partial\udelta\phi^i)(\partial\usigma\partial\urho\phi^j)\delta_{ij} &= 
\frac{1}{2}
\big(
\partial\urho\partial\usigma g_{\beta\delta} 
-\partial\urho\partial\ubeta g_{\sigma\beta}\\
&\quad
-\partial\udelta\partial\usigma g_{\beta\rho}
+\partial\udelta\partial\ubeta g_{\rho\sigma}
\big).
\end{split}
\end{align}
Hence we have
\begin{align}\label{eq: relation1}
\begin{split}
G\orhosigma g^{\beta\delta}(\partial\urho\partial\ubeta\phi^i)(\partial\usigma\partial\udelta\phi^j)\delta_{ij}
-g^{\beta\delta}(\partial\ubeta\partial\udelta \phi^i)G\orhosigma(\partial\urho\partial\usigma \phi^j)\delta_{ij}=&
\\
G\orhosigma g^{\beta\delta}(\partial\urho\partial\ubeta\phi^i)(\partial\usigma\partial\udelta\phi^j)\delta_{ij}
-\eps g^{\beta\delta}(\partial\ubeta\partial\udelta \phi^i)
\left(
\Delta_{\Gtilde} \phi^j
+\widetilde{\Gamma}\omu(\partial\umu\phi^j)
\right)
\delta_{ij}=&
\\
\frac{1}{2}\big(
G\omunu(g\partial\umu\partial\unu g\hme)-2G\orhosigma g^{\delta\beta}\partial\urho\partial\ubeta g_{\sigma\delta}
+g\omunu G\orhosigma \partial\umu\partial\unu g_{\rho\sigma}
\big).
\end{split}
\end{align}
A simple relation is also
\begin{align}
(\partial\unu\phi^i)(\partial\umu\partial_{\lambda}\phi^j)\delta_{ij}\theta^{\mu\nu} &= (\partial\umu g_{\nu\lambda})\theta\omunu.
\end{align}

\paragraph{Computation of $\tr \left(G\umunu a\omu a\onu \right)$.}
\begin{align}
\begin{split}
\tr \left(G\umunu a\omu a\onu \right) &=
\tr \left[
\widetilde{\gamma}\ualpha\widetilde{\gamma}\ubeta \theta^{\rho\alpha}(\partial\urho \theta^{\mu\beta} +
\widetilde{\gamma}\ualpha(\partial\urho \widetilde{\gamma}\ubeta)\theta^{\rho\alpha}\theta^{\mu\beta}
\right]\times \\
&\quad
\left[
\widetilde{\gamma}\ugamma\widetilde{\gamma}\udelta\theta^{\sigma\gamma}(\partial\usigma\theta^{\nu\delta}) +
\widetilde{\gamma}\ugamma(\partial\usigma \widetilde{\gamma}\udelta)\theta^{\sigma\gamma}\theta^{\nu\delta}
\right]G\umunu \\
&=
\tr \Big[
\widetilde{\gamma}\ualpha\widetilde{\gamma}\ubeta\widetilde{\gamma}\ugamma\widetilde{\gamma}\udelta
\theta^{\rho\alpha}\theta^{\sigma\gamma}(\partial\urho \theta^{\mu\beta})(\partial\usigma \theta^{\nu\delta}) G\umunu \\
&\quad
+2 \widetilde{\gamma}\ualpha\widetilde{\gamma}\ubeta\widetilde{\gamma}\ugamma(\partial\usigma \widetilde{\gamma}\udelta) 
\theta^{\rho\alpha}(\partial\urho \theta^{\mu\beta})\theta^{\sigma\gamma}\theta^{\nu\delta}G\umunu \\
&\quad
+\widetilde{\gamma}\ualpha(\partial\urho \widetilde{\gamma}\ubeta)\widetilde{\gamma}\ugamma(\partial\usigma\widetilde{\gamma}\udelta)
\theta^{\rho\alpha}\theta^{\sigma\gamma}g^{\beta\delta}
\Big]
\end{split}
\end{align}
We evaluate the trace of the Gamma matrices $\widetilde{\gamma}$ that appear in the above expression,
\begin{align}
\begin{split}
\tr \gammat\urho\gammat\usigma\gammat\ualpha\gammat\ubeta &=
k\left( g\urhosigma g_{\alpha\beta} - g_{\rho\alpha}g_{\sigma\beta} + g_{\rho\beta}g_{\sigma\alpha}\right), \\
\tr \gammat\urho\gamma_{3+j}\gammat\ualpha \gammat\ubeta &=
k\,\left((\partial\urho\phi^i)g_{\alpha\beta}-(\partial\ubeta\phi^i)g_{\rho\alpha}
+(\partial\ualpha\phi^i)g_{\rho\beta}\right) \delta_{ij}, \\
\tr \gammat\urho\gammat\usigma\gammat\ualpha \gamma_{3+i} &=
k \,\left((\partial\ualpha\phi^j)g_{\rho\sigma}-(\partial\usigma\phi^j)g_{\alpha\rho}+(\partial\urho\phi^j)g_{\alpha\sigma}\right)\delta_{ij}, \\
\tr \gammat\urho\gamma_{3+j}\gammat\ualpha\gamma_{3+i} &=
k\left(-\delta_{ij}g_{\rho\alpha}+(\delta_{kj}\delta_{li}+\delta_{ki}\delta_{jl})(\partial\urho\phi^k)(\partial\ualpha \phi^l) \right).
\end{split}
\end{align}
Here $k$ is the rank of the representation of the $\gamma$-matrices, depending on the number of extra dimensions. 
\begin{align}
\begin{split}
\widetilde{\gamma}\ualpha\widetilde{\gamma}\ubeta\widetilde{\gamma}\ugamma\widetilde{\gamma}\udelta
\theta^{\rho\alpha}\theta^{\sigma\gamma}(\partial\urho \theta^{\mu\beta})(\partial\usigma \theta^{\nu\delta}) G\umunu
&=
k\,(g\ualphabeta g_{\gamma\delta}-g_{\alpha\gamma}g_{\beta\delta}+g_{\alpha\delta}g_{\beta\gamma}) \times\\
&\quad
\theta^{\rho\alpha}\theta^{\sigma\gamma}(\partial\urho \theta^{\mu\beta})(\partial\usigma \theta^{\nu\delta})G\umunu \\
&=
k\Big\{\,G\omunu (\partial\umu \thetainv_{\nu\alpha})G\orhosigma(\partial\urho\thetainv_{\sigma\beta})g\oalphabeta \\
&\quad
-G\omunu G\orhosigma (\partial\umu \thetainv_{\rho\alpha})(\partial\unu \thetainv_{\sigma\beta})g\oalphabeta \\
&\quad
+G^{\rho\mu}G^{\sigma\nu}(\partial\urho \thetainv_{\nu\alpha})(\partial\usigma \thetainv_{\mu\beta})g\oalphabeta 
\Big\}
\end{split}
\end{align}

\begin{align}
\begin{split}
 \widetilde{\gamma}\ualpha\widetilde{\gamma}\ubeta\widetilde{\gamma}\ugamma(\partial\usigma \widetilde{\gamma}\udelta) 
\theta^{\rho\alpha}(\partial\urho \theta^{\mu\beta})\theta^{\sigma\gamma}\theta^{\nu\delta}G\umunu
&=
\tr \big[
\widetilde{\gamma}\ualpha\widetilde{\gamma}\ubeta\widetilde{\gamma}\ugamma(\partial\usigma\widetilde{\gamma}\udelta)
\theta^{\rho\alpha}(\partial\urho \theta^{\mu\beta})\theta^{\sigma\gamma}\theta^{\nu\delta}G\umunu 
\big]\\
&=k\delta_{ij}\big[
(\partial\ualpha\phi^j)g_{\beta\gamma}-(\partial\ubeta\phi^j)g_{\alpha\gamma}+(\partial\ugamma\phi^j)g\ualphabeta
\big]  \times \\
&\quad
(\partial\usigma\partial\udelta\phi^i)
\theta^{\rho\alpha}(\partial\urho \theta^{\mu\beta})\theta^{\sigma\gamma}\theta^{\nu\beta}G\umunu \\
&=
\frac{k}{2}\;\big[
(\partial\usigma g_{\alpha\delta}+\partial\udelta g_{\alpha\sigma}-\partial\ualpha g_{\sigma\delta})
g_{\beta\gamma}\theta^{\rho\alpha}(\partial\urho \theta^{\mu\beta})\theta^{\sigma\gamma}\theta^{\nu\delta}G\umunu \\
&\quad
-(\partial\usigma g_{\beta\delta}+\partial\udelta g_{\beta\sigma}-\partial\ubeta g_{\sigma\delta})
G\orhosigma(\partial\urho \theta^{\mu\beta})\theta^{\nu\delta}G\umunu \\
&\quad
+(\partial\usigma g_{\gamma\delta}+\partial\udelta g_{\gamma\sigma}-\partial\ugamma g_{\sigma\delta})
g\ualphabeta \theta^{\rho\alpha}(\partial\urho \theta^{\mu\beta})\theta^{\sigma\gamma}\theta^{\nu\delta}G\umunu
\big]\\
&=\frac{k}{2}\;\big[
G^{\mu\sigma}\theta^{\rho\alpha}(\partial\usigma g_{\alpha\delta})(\partial\urho\thetainv\umunu)g^{\nu\delta}
+G^{\mu\sigma}\theta^{\rho\alpha}(\partial\udelta g_{\alpha\sigma})(\partial\urho \thetainv\umunu)g^{\nu\delta}\\
&\quad
-G^{\mu\sigma}\theta^{\rho\alpha}(\partial\ualpha g_{\sigma\delta})(\partial\urho \thetainv\umunu)g^{\nu\delta}
-G\orhosigma\theta^{\mu\beta}(\partial\usigma g_{\beta\delta})(\partial\urho \thetainv\umunu)g^{\nu\delta}\\
&\quad
-G\orhosigma\theta^{\mu\beta}(\partial\udelta g_{\beta\sigma})(\partial\urho \thetainv\umunu)g^{\nu\delta}
+G\orhosigma\theta^{\mu\beta}(\partial\ubeta g_{\sigma\delta})(\partial\urho \thetainv\umunu)g^{\nu\delta} \\
&\quad
-2 G^{\rho\mu}(\partial\urho \thetainv\umunu)G^{\sigma\lambda}(\partial\usigma \thetainv_{\lambda\delta})g^{\nu\delta}
\big] 
\end{split}
\end{align}

\begin{align}
\begin{split}
\widetilde{\gamma}\ualpha(\partial\urho \widetilde{\gamma}\ubeta)\widetilde{\gamma}\ugamma(\partial\usigma\widetilde{\gamma}\udelta)
\theta^{\rho\alpha}\widetilde{\gamma}^{\sigma\gamma}g^{\beta\delta}
&=
\tr\big[
\widetilde{\gamma}\ualpha(\partial\urho\widetilde{\gamma}\ubeta)\widetilde{\gamma}\ugamma(\partial\usigma\widetilde{\gamma}\udelta)
\theta^{\rho\alpha}\theta^{\sigma\gamma}g^{\beta\delta} 
\big]\\
&=
k\;\big[
-\delta_{ij}g_{\alpha\gamma}+(\delta_{ki}\delta_{lj}+\delta_{kj}\delta_{li})(\partial\ualpha\phi^k)(\partial\ugamma\phi^l)
\big] \times \\
&\quad
(\partial\urho\partial\ubeta\phi^i)(\partial\usigma\partial\udelta\phi^j)\theta^{\rho\alpha}\theta^{\sigma\gamma}g^{\beta\delta} \\
&=
k\;\big[
-(\partial\urho\partial\ubeta \phi^i)(\partial\usigma\partial\udelta\phi^j)\delta_{ij}G\orhosigma g^{\beta\delta} 
+(\partial\urho g\ualphabeta)\theta^{\rho\alpha}(\partial\usigma g_{\gamma\delta})\theta^{\sigma\gamma}g^{\beta\delta} \\
&\quad
+\frac{1}{4}(\partial\usigma g_{\alpha\delta}+\partial\udelta g_{\alpha\sigma}-\partial\ualpha g_{\sigma\delta})
(\partial\urho g_{\gamma\beta}+\partial\ubeta g_{\gamma\rho}-\partial\ugamma g_{\rho\beta})
\theta^{\rho\alpha}\theta^{\sigma\gamma}g^{\beta\delta}
\big]\\
&=
k\;\big[
-\frac{1}{2}\big(
G\omunu(g\partial\umu\partial\unu g\hme)-2G\orhosigma g^{\delta\beta}\partial\urho\partial\ubeta g_{\sigma\delta}
+g\omunu G\orhosigma \partial\umu\partial\unu g_{\rho\sigma}
\big)\\
&\quad
+G\omunu(\partial\umu\thetainv_{\nu\alpha})G\orhosigma(\partial\urho\thetainv_{\sigma\beta})g\oalphabeta \\
&\quad
+\frac{1}{2}\theta^{\rho\alpha}(\partial\urho g_{\gamma\beta})\theta^{\sigma\gamma}(\partial\usigma g_{\alpha\delta})g^{\delta\beta} \\
&\quad
+\frac{1}{2}\theta\omunu\theta\orhosigma(\partial\umu g_{\rho\alpha})(\partial\unu g_{\sigma\beta})g\oalphabeta \\
&\quad
-\frac{1}{4}\theta\omunu\theta\orhosigma(\partial\ualpha g_{\mu\rho})(\partial\ubeta g_{\nu\sigma})g\oalphabeta
 \\
&\quad
-\eps g^{\beta\delta}(\partial\ubeta\partial\udelta \phi^i)
\left(
\Delta_{\Gtilde} \phi^j
+\widetilde{\Gamma}\omunu(\partial\umu\phi^j)
\right)
\delta_{ij}
\big]
\end{split}
\end{align}
In the last step we have used Eq.(\ref{eq: relation1}). The explicit expression for the whole term  is then
\begin{align}
\begin{split}
- \frac{\ems}{4}\tr\left(G\omunu a\umu a\unu\right) 
&=
- k\frac{\ems}{4}\Big\{
G\omunu (\partial\umu \thetainv_{\nu\alpha})G\orhosigma(\partial\urho\thetainv_{\sigma\beta})g\oalphabeta  \\
&\quad
-G\omunu G\orhosigma (\partial\umu \thetainv_{\rho\alpha})(\partial\unu \thetainv_{\sigma\beta})g\oalphabeta
+G^{\rho\mu}G^{\sigma\nu}(\partial\urho \thetainv_{\nu\alpha})(\partial\usigma \thetainv_{\mu\beta})g\oalphabeta \\
&\quad
+G^{\mu\sigma}\theta^{\rho\alpha}(\partial\usigma g_{\alpha\delta})(\partial\urho\thetainv\umunu)g^{\nu\delta}
+G^{\mu\sigma}\theta^{\rho\alpha}(\partial\udelta g_{\alpha\sigma})(\partial\urho \thetainv\umunu)g^{\nu\delta}\\
&\quad
-G^{\mu\sigma}\theta^{\rho\alpha}(\partial\ualpha g_{\sigma\delta})(\partial\urho \thetainv\umunu)g^{\nu\delta}
-G\orhosigma\theta^{\mu\beta}(\partial\usigma g_{\beta\delta})(\partial\urho \thetainv\umunu)g^{\nu\delta}\\
&\quad
-G\orhosigma\theta^{\mu\beta}(\partial\udelta g_{\beta\sigma})(\partial\urho \thetainv\umunu)g^{\nu\delta}
+G\orhosigma\theta^{\mu\beta}(\partial\ubeta g_{\sigma\delta})(\partial\urho \thetainv\umunu)g^{\nu\delta} \\
&\quad
-2 G^{\rho\mu}(\partial\urho \thetainv\umunu)G^{\sigma\lambda}(\partial\usigma \thetainv_{\lambda\delta})g^{\nu\delta} \\
&\quad
-\frac{1}{2}\big(
G\omunu(g\partial\umu\partial\unu g\hme)-2G\orhosigma g^{\delta\beta}\partial\urho\partial\ubeta g_{\sigma\delta}
+g\omunu G\orhosigma \partial\umu\partial\unu g_{\rho\sigma}
\big)\\
&\quad
+G\omunu(\partial\umu\thetainv_{\nu\alpha})G\orhosigma(\partial\urho\thetainv_{\sigma\beta})g\oalphabeta 
+\frac{1}{2}\theta^{\rho\alpha}(\partial\urho g_{\gamma\beta})\theta^{\sigma\gamma}(\partial\usigma g_{\alpha\delta})g^{\delta\beta} \\
&\quad
+\frac{1}{2}\theta\omunu\theta\orhosigma(\partial\umu g_{\rho\alpha})(\partial\unu g_{\sigma\beta})g\oalphabeta 
-\frac{1}{4}\theta\omunu\theta\orhosigma(\partial\ualpha g_{\mu\rho})(\partial\ubeta g_{\nu\sigma})g\oalphabeta
\Big\}\\
&\quad
+\frac{k}{4}g\omunu(\partial\umu\partial\unu \phi^i)
\left(
\Delta_{\Gtilde}\phi^j+\widetilde{\Gamma}\omu\partial\umu \phi^j
\right)\delta_{ij}.
\end{split}
\end{align}
\paragraph{Computation of $\tr(-G\omunu(\partial\umu G_{\nu\rho})a\orho)$ and $\tr(\partial\umu a\omu$).}
Next we deal with the remaining two terms in $\tr\cE$. They turn out to cancel each other and moreover both are zero for on-shell geometries.  
We evaluate again the trace.
\begin{align}
\begin{split}
\tr (\partial\umu \widetilde{\gamma}\ualpha)\widetilde{\gamma}\ubeta &=
\tr \big[
\gamma_{3+i}(\partial\umu\partial\ualpha \phi^i)(\gamma\ubeta + \gamma_{3+j}(\partial\ubeta \phi^j)) 
\big]\\
&=
k\delta_{ij}(\partial\umu\partial\ualpha\phi^i)(\partial\ubeta\phi^j)\\
&=
\frac{k}{2}(\partial\umu g_{\alpha\beta}+\partial\ualpha g_{\mu\beta} -\partial\ubeta g_{\mu\alpha}) 
\end{split}
\end{align}
\begin{align}
\begin{split}
\tr (\partial\umu\widetilde{\gamma}\ualpha)(\partial\unu \widetilde{\gamma}\ubeta)\theta^{\nu\alpha}\theta^{\mu\beta}
&=\Tr\gamma_{3+i}\gamma_{3+j}(\partial\umu\partial\ualpha\phi^i)(\partial\unu\partial\ubeta \phi^j)\theta^{\nu\alpha}\theta^{\mu\beta}\\
&=k\delta_{ij}(\partial\umu\partial\ualpha\phi^i)(\partial\unu\partial\ubeta \phi^j)\theta^{\nu\alpha}\theta^{\mu\beta}\\
&=\frac{1}{2}(\partial\umu\partial\unu g\ualphabeta+\partial\umu\partial\ubeta g_{\alpha\nu}-\partial\umu\partial\ualpha g_{\nu\beta})
\theta^{\nu\alpha}\theta^{\mu\beta}\\
&=
\theta^{\mu\alpha}\theta^{\nu\beta}\partial\umu\partial\unu g\ualphabeta
\end{split}
\end{align}
First we consider the computation of $-G\omunu(\partial\umu G_{\nu\rho})a\orho$.
\begin{align}\label{eq: G partial G a}
\begin{split}
\tr a\orho &= \Tr\big[
\widetilde{\gamma}\ualpha\widetilde{\gamma}\ubeta \theta^{\sigma\alpha}(\partial\usigma \theta^{\rho\beta})
+\widetilde{\gamma}\ualpha(\partial\usigma\widetilde{\gamma}\ubeta)\theta^{\sigma\alpha}\theta^{\rho\beta}
\big] \\
&=
k \big[
g\ualphabeta \theta^{\sigma\alpha}(\partial\usigma \theta^{\rho\beta})+
\frac{1}{2}(\partial\usigma g\ualphabeta +\partial\ubeta g_{\sigma\alpha}-\partial\ualpha g_{\sigma\beta})\theta^{\sigma\alpha}\theta^{\rho\beta}
\big]\\
&=
k\big[
\theta^{\sigma\alpha}(\partial\usigma\theta^{\rho\beta})g\ualphabeta 
+\theta^{\sigma\alpha}\theta^{\rho\alpha}(\partial\usigma g\ualphabeta)
\big]\\
&=
-k\,\eps \widetilde{\Gamma}\orho.
\end{split}
\end{align}
The remaining term $\tr \partial\umu a\omu$ gives
\begin{align}\label{eq: partial a}
\begin{split}
\tr \partial\umu a\omu &= \Tr \big[
(\partial\umu\widetilde{\gamma}\ualpha)\widetilde{\gamma}\ubeta\theta^{\nu\alpha}(\partial\unu\theta^{\mu\beta})
+\widetilde{\gamma}\ualpha(\partial\umu\widetilde{\gamma}\ubeta)\theta^{\nu\alpha}(\partial\unu\theta^{\mu\beta})\\
&\quad
+\widetilde{\gamma}\ualpha\widetilde{\gamma}\ubeta(\partial\umu\theta^{\nu\alpha})(\partial\unu\theta^{\mu\beta})
+\widetilde{\gamma}\ualpha\widetilde{\gamma}\ubeta\theta^{\nu\beta}\partial\unu\partial\umu\theta^{\mu\beta}\\
&\quad
+(\partial\umu\widetilde{\gamma}\ualpha)(\partial\unu \widetilde{\gamma}\ubeta)\theta^{\nu\alpha}\theta^{\mu\beta}
+\widetilde{\gamma}\ualpha(\partial\unu\widetilde{\gamma}\ubeta)(\partial\umu\theta^{\nu\alpha})\theta^{\mu\beta}\\
&\quad
+\widetilde{\gamma}\ualpha(\partial\unu\widetilde{\gamma}\ubeta)\theta^{\nu\alpha}(\partial\umu\theta^{\mu\beta})
\big]\\
&=
\frac{k}{2}\big\{
(\partial\umu g\ualphabeta +\partial\ualpha g_{\mu\beta}-\partial\ubeta g_{\mu\alpha})\theta^{\nu\beta}(\partial\unu \theta^{\mu\beta})\\
&\quad
+(\partial\umu g\ualphabeta +\partial\ubeta g_{\mu\alpha}-\partial\ualpha g_{\mu\beta})\theta^{\nu\beta}(\partial\unu \theta^{\mu\beta})\\
&\quad
+2(\partial\umu\theta^{\nu\alpha})(\partial\unu\theta^{\mu\beta})g\ualphabeta
+2\theta^{\mu\alpha}(\partial\umu\partial\unu\theta^{\nu\beta})g\ualphabeta
+2\theta^{\mu\alpha}\theta^{\nu\beta}\partial\umu\partial\unu g\ualphabeta\\
&\quad
+(\partial\unu g\ualphabeta +\partial\ubeta g_{\alpha\nu}-\partial\ualpha g_{\nu\beta})(\partial\umu\theta^{\nu\alpha})\theta^{\mu\beta}\\
&\quad
+(\partial\unu g\ualphabeta +\partial\ubeta g_{\alpha\nu}-\partial\ualpha g_{\nu\beta})\theta^{\nu\alpha}(\partial\umu\theta^{\mu\beta})
\big\}\\
&=
k\,\big\{
2\theta^{\mu\alpha}(\partial\umu\theta^{\nu\beta})(\partial\unu g\ualphabeta)
+\theta^{\mu\alpha}(\partial\unu\theta^{\nu\beta})(\partial\umu g\ualphabeta)\\
&\quad
+(\partial\umu\theta^{\nu\alpha})(\partial\unu\theta^{\mu\beta})g\ualphabeta
+\theta^{\mu\alpha}(\partial\umu\partial\unu\theta^{\nu\beta})g\ualphabeta
+\theta^{\mu\alpha}\theta^{\nu\beta}\partial\umu\partial\unu g\ualphabeta
\big\} \\
&=
k\,
\partial\umu \big\{
\theta^{\nu\alpha}(\partial\unu\theta^{\mu\beta})g\ualphabeta
+\theta^{\nu\alpha}\theta^{\mu\beta}(\partial\unu g\ualphabeta)
\big\} \\
&=-k\,\partial\umu \left(
\eps \widetilde{\Gamma}\omu
\right).
\end{split}
\end{align}
Due to Eq. (\ref{eq: G partial G a}) and (\ref{eq: partial a}) we find
\begin{align}
\begin{split}
\tr\,\left(\Gtilde\omunu \partial\umu 
\big(
\Gtilde_{\nu\rho}\widetilde{a}\orho + \Gtilde_{\nu\rho}\widetilde{\Gamma}\orho
\big)\right)=0
\end{split}
\end{align}
and thus
\begin{align}\label{eq: trE final}
\begin{split}
\tr \cE &= - \tr \left(
\frac{1}{4}\Gtilde\umunu \widetilde{a}\omu \widetilde{a}\onu
-\frac{1}{4}\Gtilde\umunu \widetilde{\Gamma}\omu \widetilde{\Gamma}\onu
\right) \\
&=- k\frac{\ems}{4}\Big\{
G\omunu (\partial\umu \thetainv_{\nu\alpha})G\orhosigma(\partial\urho\thetainv_{\sigma\beta})g\oalphabeta  \\
&\quad
-G\omunu G\orhosigma (\partial\umu \thetainv_{\rho\alpha})(\partial\unu \thetainv_{\sigma\beta})g\oalphabeta
+G^{\rho\mu}G^{\sigma\nu}(\partial\urho \thetainv_{\nu\alpha})(\partial\usigma \thetainv_{\mu\beta})g\oalphabeta \\
&\quad
+G^{\mu\sigma}\theta^{\rho\alpha}(\partial\usigma g_{\alpha\delta})(\partial\urho\thetainv\umunu)g^{\nu\delta}
+G^{\mu\sigma}\theta^{\rho\alpha}(\partial\udelta g_{\alpha\sigma})(\partial\urho \thetainv\umunu)g^{\nu\delta}\\
&\quad
-G^{\mu\sigma}\theta^{\rho\alpha}(\partial\ualpha g_{\sigma\delta})(\partial\urho \thetainv\umunu)g^{\nu\delta}
-G\orhosigma\theta^{\mu\beta}(\partial\usigma g_{\beta\delta})(\partial\urho \thetainv\umunu)g^{\nu\delta}\\
&\quad
-G\orhosigma\theta^{\mu\beta}(\partial\udelta g_{\beta\sigma})(\partial\urho \thetainv\umunu)g^{\nu\delta}
+G\orhosigma\theta^{\mu\beta}(\partial\ubeta g_{\sigma\delta})(\partial\urho \thetainv\umunu)g^{\nu\delta} \\
&\quad
-2 G^{\rho\mu}(\partial\urho \thetainv\umunu)G^{\sigma\lambda}(\partial\usigma \thetainv_{\lambda\delta})g^{\nu\delta} \\
&\quad
-\frac{1}{2}\big(
G\omunu(g\partial\umu\partial\unu g\hme)-2G\orhosigma g^{\delta\beta}\partial\urho\partial\ubeta g_{\sigma\delta}
+g\omunu G\orhosigma \partial\umu\partial\unu g_{\rho\sigma}
\big)\\
&\quad
+G\omunu(\partial\umu\thetainv_{\nu\alpha})G\orhosigma(\partial\urho\thetainv_{\sigma\beta})g\oalphabeta 
+\frac{1}{2}\theta^{\rho\alpha}(\partial\urho g_{\gamma\beta})\theta^{\sigma\gamma}(\partial\usigma g_{\alpha\delta})g^{\delta\beta} \\
&\quad
+\frac{1}{2}\theta\omunu\theta\orhosigma(\partial\umu g_{\rho\alpha})(\partial\unu g_{\sigma\beta})g\oalphabeta 
-\frac{1}{4}\theta\omunu\theta\orhosigma(\partial\ualpha g_{\mu\rho})(\partial\ubeta g_{\nu\sigma})g\oalphabeta
\Big\}\\
&\quad
+\frac{k}{4}g\omunu(\partial\umu\partial\unu \phi^i)
\left(
\Delta_{\Gtilde}\phi^j+\widetilde{\Gamma}\omu\partial\umu \phi^j
\right)\delta_{ij} \\
&\quad
+\frac{k}{4}\Gtilde\umunu \widetilde{\Gamma}\omu \widetilde{\Gamma}\onu.
\end{split}
\end{align}
That means also that for on-shell geometries $\tr\cE$ is given solely  by 
\begin{align}
\tr\cE = -\frac{e^{-\sigma}}{4} \tr\, G\omunu a\umu a\unu. 
\end{align}

\section{Appendix B: Covariance of $\tr\cE$}

We aim to show that  $\tr \cE$ can be written in covariant manner. If so, we can change to a normal coordinate system, which will simplify 
$\tr \cE$ and the Ricci scalar enormously. However, notice that now
$\tr \cE$ should be related to the Ricci scalar directly and not only
under the integral, where we would be allowed to use partial
integration. 
Since normal coordinates make sense only at a point, 
partial integration is not admissible here. 
\paragraph{Notation.} We  distinguish between the \emph{effective metric} $\Gtilde\umunu$ and the \emph{background metric} $g\umunu$. The covariant derivatives and Christoffel symbols  with respect to the background metric $g\umunu$ as  important in this section are written as $\nabla\umu$ and $\Gamma\omu\urhosigma$.

By using expressions containing derivatives of $g\umunu$,
\begin{align}
\begin{split}
\partial\ulambda g\umunu &= (\partial\ulambda \partial\umu \phi^i)(\partial\unu \phi^j)\delta_{ij} 
+(\partial\umu\phi^i)(\partial\ulambda\partial\unu \phi^j)\delta_{ij},\\
\partial\unu g_{\lambda\mu} &= (\partial\unu \partial\ulambda \phi^i)(\partial\umu \phi^j)\delta_{ij}
+(\partial\ulambda\phi^i)(\partial\unu\partial\umu \phi^j)\delta_{ij}, \\
\partial\umu g_{\nu\lambda} &= (\partial\umu \partial\unu \phi^i)(\partial\ulambda \phi^j)\delta_{ij} 
+(\partial\unu\phi^i)(\partial\umu\partial\ulambda \phi^j)\delta_{ij},
\end{split}
\end{align}
we find the following relation
\begin{align}\label{relation phi Gamma}
\begin{split}
(\partial_{\lambda}\phi^i)(\partial\unu\partial\umu\phi^j)\delta_{ij} &=
\frac{1}{2}\Big( \partial\umu g_{\nu\lambda}+\partial\unu g_{\lambda\mu} - \partial_{\lambda} g\umunu \Big) \\
&=
\Gamma\osigma_{\mu\nu}g_{\sigma\lambda}.
\end{split}
\end{align}
Now we are able to rewrite $\tr \cE$ in terms of the Christoffel symbols $\Gamma\umunu\orho[g]$. 
\begin{align}\label{eq: tr e}
\begin{split}
\tr \cE &=
-\frac{\ems}{4}\,\tr \left(a\omu a\onu G\umunu\right)
+\frac{1}{4}\,\tr \left(\Gtilde\umunu \widetilde{\Gamma}\omu \widetilde{\Gamma}\onu \right)\\
&=-\frac{\ems}{4}\,
\tr \Big[
\widetilde{\gamma}\ualpha\widetilde{\gamma}\ubeta\widetilde{\gamma}\ugamma\widetilde{\gamma}\udelta
\theta^{\rho\alpha}\theta^{\sigma\gamma}(\partial\urho \theta^{\mu\beta})(\partial\usigma \theta^{\nu\delta}) G\umunu \\
&\quad
+2 \widetilde{\gamma}\ualpha\widetilde{\gamma}\ubeta\widetilde{\gamma}\ugamma(\partial\usigma \widetilde{\gamma}\udelta) 
\theta^{\rho\alpha}(\partial\urho \theta^{\mu\beta})\theta^{\sigma\gamma}\theta^{\nu\delta}G\umunu \\
&\quad
+\widetilde{\gamma}\ualpha(\partial\urho \widetilde{\gamma}\ubeta)\widetilde{\gamma}\ugamma(\partial\usigma\widetilde{\gamma}\udelta)
\theta^{\rho\alpha}\theta^{\sigma\gamma}g^{\beta\delta}
\Big]\\
&\quad
+\frac{\ems}{4}\,\tr \left(\Gtilde\umunu \widetilde{\Gamma}\omu \widetilde{\Gamma}\onu\right) \\
&=-\frac{\ems}{4}\,k\,
\Big[
(g\ualphabeta g_{\gamma\delta}-g_{\alpha\gamma}g_{\beta\delta}+g_{\alpha\delta}g_{\beta\gamma})
\theta^{\rho\alpha}\theta^{\sigma\gamma}(\partial\urho \theta^{\mu\beta})(\partial\usigma \theta^{\nu\delta}) G\umunu \\
&\quad
+2\delta_{ij}\big[
(\partial\ualpha\phi^j)g_{\beta\gamma}-(\partial\ubeta\phi^j)g_{\alpha\gamma}+(\partial\ugamma\phi^j)g\ualphabeta
\big] (\partial\usigma\partial\udelta\phi^i)
\theta^{\rho\alpha}(\partial\urho \theta^{\mu\beta})\theta^{\sigma\gamma}\theta^{\nu\delta}G\umunu \\
&\quad
+\big[
-\delta_{ij}g_{\alpha\gamma}+(\delta_{ki}\delta_{lj}+\delta_{kj}\delta_{li})(\partial\ualpha\phi^k)(\partial\ugamma\phi^l)
\big] 
(\partial\urho\partial\ubeta\phi^i)(\partial\usigma\partial\udelta\phi^j)\theta^{\rho\alpha}\theta^{\sigma\gamma}g^{\beta\delta}
\Big]
\\
&\quad
+\frac{1}{4}\,\tr \left(\Gtilde\umunu \widetilde{\Gamma}\omu \widetilde{\Gamma}\onu\right).
\end{split}
\end{align}
Since the first term in Eq. (\ref{eq: tr e}) does not contain a partial derivative of $g\umunu$, we begin with the second term,
\begin{align}
\begin{split}
2\,\tr\left(
\widetilde{\gamma}\ualpha\widetilde{\gamma}\ubeta\widetilde{\gamma}\ugamma(\partial\usigma\widetilde{\gamma}\urho)
\theta^{\rho\alpha}(\partial\urho\theta^{\mu\beta})\theta^{\sigma\gamma}\theta^{\nu\delta}G\umunu
\right)
&=2\,k\,\delta_{ij}\,\Big(
(\partial\ualpha\phi^i)g_{\beta\gamma}-(\partial\ubeta\phi^j)g_{\alpha\gamma}
+(\partial\ugamma\phi^j)g\ualphabeta
\Big)\times \\
&\quad (\partial\usigma\partial\udelta\phi^i)\theta^{\rho\alpha}
(\partial\urho\theta^{\mu\beta})\theta^{\sigma\gamma}\theta^{\nu\delta}G\umunu \\
&=2\,k\,\Big(
g_{\alpha\lambda}\Gamma^{\lambda}_{\sigma\delta}g_{\beta\delta}\theta^{\rho\alpha}(\partial\urho\theta^{\mu\beta})
\theta^{\sigma\gamma}\theta^{\nu\delta}G\umunu \\
&\quad
-g_{\beta\lambda}\Gamma^{\lambda}_{\sigma\delta}g_{\alpha\gamma}\theta^{\rho\alpha}(\partial\urho\theta^{\mu\beta})
\theta^{\sigma\gamma}\theta^{\nu\delta}G\umunu \\
&\quad
+g_{\gamma\lambda}\Gamma^{\lambda}_{\sigma\delta}g\ualphabeta\theta^{\rho\alpha}(\partial\urho\theta^{\mu\beta})
\theta^{\sigma\gamma}\theta^{\nu\delta}G\umunu 
\Big)\\
&=
2\,k\,\Big(
\underbrace{g_{\alpha\lambda}\Gamma^{\lambda}_{\sigma\delta}G^{\mu\sigma}\theta^{\rho\alpha}(\partial\urho\thetainv\umunu)g^{\nu\delta}}_{(a)}\\
&\quad
-\underbrace{g_{\alpha\lambda}\Gamma^{\lambda}_{\sigma\delta}G\orhosigma(\partial\urho\thetainv\umunu)\theta^{\mu\alpha}g^{\nu\delta}}_{(b)}
\\
&\quad
+\underbrace{g_{\alpha\lambda}\Gamma^{\lambda}_{\sigma\delta}G^{\rho\mu}(\partial\urho\thetainv_{\mu\nu})\theta^{\sigma\alpha}g^{\nu\delta}}_{(c)}
\Big).
\end{split}
\end{align}
Next we address the third term in Eq.(\ref{eq: tr e}). We write
\begin{align}
\begin{split}
(\partial\urho\partial\ubeta\phi^i)(\partial\usigma\partial\udelta\phi^j)\delta_{ij}
+(\partial\ubeta\phi^i)(\partial\urho\partial\usigma\partial\udelta\phi^j)\delta_{ij}
=(\partial\urho g_{\beta\lambda})\Gamma\olambda_{\sigma\delta}
+g_{\beta\lambda}(\partial\urho \Gamma\olambda_{\sigma\delta}).
\end{split}
\end{align}
and subtract from this equation the same equation, interchanging this time the indices $\rho$ and $\delta$. This gives
\begin{align}
\begin{split}
&(\partial\urho\partial\ubeta\phi^i)(\partial\usigma\partial\udelta\phi^j)\delta_{ij}G\orhosigma g^{\delta\beta}
-g^{\delta\beta}(\partial\udelta\partial\ubeta\phi^i)G\orhosigma(\partial\urho\partial\usigma\phi^j)\delta_{ij}=
\\
&(\partial\urho\partial\ubeta\phi^i)(\partial\usigma\partial\udelta\phi^j)\delta_{ij}G\orhosigma g^{\delta\beta}
-\eps g^{\mu\nu}(\partial\umu\partial\unu\phi^i)
\left(
\Delta_{\Gtilde}\phi^i+\widetilde{\Gamma}\orho \partial\urho \phi^j
\right)
\delta_{ij}=
\\
&G\orhosigma g^{\delta\beta}(\partial\urho g_{\beta\lambda})\Gamma\olambda_{\sigma\delta} 
+G\orhosigma(\partial\urho \Gamma\olambda_{\sigma\lambda}) 
-G\orhosigma g^{\beta\delta}(\partial\udelta g_{\beta\lambda})\Gamma\olambda\urhosigma
-G\orhosigma (\partial\ulambda \Gamma\olambda\urhosigma).
\end{split}
\end{align}
Using
\begin{align}
\begin{split}
\partial\urho g_{\beta\lambda}=\Gamma^{\eta}_{\rho\beta}g_{\eta\lambda}+\Gamma^{\eta}_{\rho\lambda}g_{\beta\eta}
\end{split}
\end{align}
one finds
\begin{align}
\begin{split}
(\partial\urho\partial\ubeta\phi^i)(\partial\usigma\partial\udelta\phi^j)G\orhosigma g^{\delta\beta}&=
G\orhosigma g^{\delta\beta}\Gamma^{\eta}_{\rho\beta}\Gamma\olambda_{\sigma\delta}g_{\eta\lambda}
+G\orhosigma\Gamma\odelta_{\rho\lambda}\Gamma\olambda_{\sigma\delta}+G\orhosigma(\partial\urho \Gamma\olambda_{\sigma\lambda}) \\
&\quad
-G\orhosigma g^{\delta\beta}\Gamma^{\eta}_{\delta\beta}\Gamma\olambda\urhosigma g_{\eta\lambda}
-G\orhosigma\Gamma^{\eta}_{\eta\lambda}\Gamma\olambda\urhosigma- G\orhosigma(\partial\ulambda \Gamma\olambda\urhosigma)\\
&\quad
+\eps g^{\mu\nu}(\partial\umu\partial\unu\phi^i)
\left(
\Delta_{\Gtilde}\phi^i+\widetilde{\Gamma}\orho \partial\urho \phi^j
\right)
\delta_{ij}\\
&=
G\orhosigma g^{\delta\beta}\Gamma^{\eta}_{\rho\beta}\Gamma\olambda_{\sigma\delta}g_{\eta\lambda}
-G\orhosigma g^{\delta\beta}\Gamma^{\eta}_{\delta\beta}\Gamma\olambda\urhosigma g_{\eta\lambda} \\
&\quad
+G\orhosigma \Big\{
\Gamma^{\delta}_{\rho\lambda}\Gamma\olambda_{\sigma\delta}
-\Gamma^{\lambda}_{\rho\sigma}\Gamma^{\delta}_{\delta\lambda} 
+\partial\urho \Gamma\olambda_{\sigma\lambda}
-\partial\ulambda \Gamma\olambda\urhosigma
\Big\}\\
&\quad
+\eps g^{\mu\nu}(\partial\umu\partial\unu\phi^i)
\left(
\Delta_{\Gtilde}\phi^i+\widetilde{\Gamma}\orho \partial\urho \phi^j
\right)
\delta_{ij}\\
&=
G\orhosigma g^{\delta\beta}\Gamma^{\eta}_{\rho\beta}\Gamma\olambda_{\sigma\delta}g_{\eta\lambda}
-G\orhosigma g^{\delta\beta}\Gamma^{\eta}_{\delta\beta}\Gamma\olambda\urhosigma g_{\eta\lambda} \\
&\quad
+\eps g^{\mu\nu}(\partial\umu\partial\unu\phi^i)
\left(
\Delta_{\Gtilde}\phi^i+\widetilde{\Gamma}\orho \partial\urho \phi^j
\right)
\delta_{ij}\\
&\quad
- G\omunu R\umunu[g].
\end{split}
\end{align}
We obtain for the third term of $\tr\cE$
\begin{align}
\begin{split}
\tr \left(
\widetilde{\gamma}\ualpha(\partial\urho\widetilde{\gamma}\ubeta)\widetilde{\gamma}\ugamma(\partial\usigma\widetilde{\gamma}\udelta)
\theta^{\rho\alpha}\theta^{\sigma\gamma}g^{\beta\delta}
\right)
&=
k\,\Big(
-\delta_{ij}g_{\alpha\gamma}
+\left(
\delta_{ki}\delta_{lj}+\delta_{kj}\delta_{li}
\right)
(\partial\ualpha\phi^k)(\partial\ugamma\phi^l)
\Big)\times\\
&\quad
(\partial\urho\partial\ubeta\phi^i)(\partial\usigma\partial\udelta\phi^j)\theta^{\rho\alpha}\theta^{\sigma\gamma}g^{\beta\delta}\\
&=
k\,\Big(
-\underbrace{G\orhosigma g^{\delta\beta}\Gamma^{\eta}_{\rho\beta}\Gamma\olambda_{\sigma\delta}g_{\eta\lambda}}_{(f)}
+G\orhosigma g^{\delta\beta}\Gamma^{\eta}_{\delta\beta}\Gamma\olambda\urhosigma g_{\eta\lambda}\\
&\quad
+ G\omunu R\umunu[g]
-\eps g^{\mu\nu}(\partial\umu\partial\unu\phi^i)
\left(
\Delta_{\Gtilde}\phi^i+\widetilde{\Gamma}\orho \partial\urho \phi^j
\right)
\delta_{ij}
\\
&\quad
+\underbrace{g_{\alpha\lambda}\Gamma^{\lambda}_{\rho\beta}g_{\gamma\lambda\oprime}\Gamma^{\lambda\oprime}_{\sigma\delta}
\theta^{\rho\alpha}\theta^{\sigma\gamma}g^{\beta\delta}}_{(d)}
+\underbrace{g_{\alpha\lambda}\Gamma^{\lambda}_{\sigma\delta}g_{\gamma\lambda\oprime}\Gamma^{\lambda\oprime}_{\rho\beta}
\theta^{\rho\alpha}\theta^{\sigma\gamma}g^{\beta\delta}}_{(e)}
\Big).
\end{split}
\end{align}
Let us write the result in an unconventional but simple way. 
\begin{align}
\begin{split}
-\frac{\ems}{4}\tr G\umunu a\omu a\onu &= -\ems\frac{k}{4}\Big\{\,G\omunu (\partial\umu \thetainv_{\nu\alpha})G\orhosigma(\partial\urho\thetainv_{\sigma\beta})g\oalphabeta 
-G\omunu G\orhosigma (\partial\umu \thetainv_{\rho\alpha})(\partial\unu \thetainv_{\sigma\beta})g\oalphabeta \\
&\quad
+G^{\rho\mu}G^{\sigma\nu}(\partial\urho \thetainv_{\nu\alpha})(\partial\usigma \thetainv_{\mu\beta})g\oalphabeta 
+\sum_{i=a}^{f}(i) + G\omunu R\umunu [g] 
+g\omunu \Gamma\olambda\umunu G\orhosigma\Gamma\urhosigma^{\eta}g_{\lambda\eta}\\
&\quad
-\eps g\omunu (\partial\umu\partial\unu \phi^i)
\left(
\Delta_{\Gtilde}\phi^j+\widetilde{\Gamma}\orho (\partial\urho\phi^j)
\right)\delta_{ij}
\Big\},
\end{split}
\end{align}
where the terms $(i), i=a\hdots f$ refer to the terms denoted via curly brace. 

Next consider the first term of Eq.(\ref{eq: tr e}),
\begin{align}\label{eq: first term}
\begin{split}
\widetilde{\gamma}\ualpha\widetilde{\gamma}\ubeta\widetilde{\gamma}\ugamma\widetilde{\gamma}\udelta
\theta^{\rho\alpha}\theta^{\sigma\gamma}(\partial\urho \theta^{\mu\beta})(\partial\usigma \theta^{\nu\delta}) G\umunu
&=
k\Big\{\,G\omunu (\partial\umu \thetainv_{\nu\alpha})G\orhosigma(\partial\urho\thetainv_{\sigma\beta})g\oalphabeta \\
&\quad
-G\omunu G\orhosigma (\partial\umu \thetainv_{\rho\alpha})(\partial\unu \thetainv_{\sigma\beta})g\oalphabeta \\
&\quad
+G^{\rho\mu}G^{\sigma\nu}(\partial\urho \thetainv_{\nu\alpha})(\partial\usigma \thetainv_{\mu\beta})g\oalphabeta 
\Big\}.
\end{split}
\end{align}
In order to write $\tr\cE$ covariantly we replace every partial derivative by a covariant derivative $\nabla\umu$. 
\begin{align}
\begin{split}
G\omunu(\nabla\umu\thetainv_{\nu\alpha})G\orhosigma(\nabla\urho\thetainv_{\sigma\beta})g\oalphabeta&=
G\omunu G\orhosigma(\partial\umu\thetainv_{\nu\alpha}-\Gamma\olambda_{\mu\nu}\thetainv_{\lambda\alpha}
-\Gamma\olambda_{\mu\alpha}\thetainv_{\nu\lambda})\times \\
&\quad
(\partial\urho\thetainv_{\sigma\beta}-\Gamma^{\lambda\oprime}\urhosigma\thetainv_{\lambda\oprime\beta}
-\Gamma^{\lambda\oprime}_{\rho\beta}\thetainv_{\sigma\lambda\oprime})g\oalphabeta \\
&=
G\omunu G\orhosigma(\partial\umu\thetainv_{\nu\alpha})(\partial\urho\thetainv_{\sigma\beta})g\oalphabeta 
-2G\omunu G\orhosigma\Gamma^{\lambda}\umunu\thetainv_{\lambda\alpha}(\partial\urho\thetainv_{\sigma\beta})g\oalphabeta \\
&\quad
+2\Gamma\olambda_{\mu\alpha}\theta\omunu g_{\nu\lambda}G\orhosigma(\partial\urho\thetainv_{\sigma\beta})g\oalphabeta
+G\omunu G\orhosigma\Gamma\umunu\olambda\Gamma^{\lambda\oprime}\urhosigma G_{\lambda\lambda\oprime}\\
&\quad
-2G\orhosigma\Gamma\olambda_{\mu\alpha}\Gamma^{\lambda\oprime}_{\rho\sigma}
\theta^{\mu\nu}g_{\nu\lambda}\thetainv_{\lambda\oprime\beta}g\oalphabeta
+\Gamma\olambda_{\mu\alpha}\Gamma^{\lambda\oprime}_{\rho\beta}\theta\omunu g_{\nu\lambda}\theta\orhosigma g_{\sigma\lambda\oprime}g\oalphabeta
\end{split}
\end{align}
\begin{align}
\begin{split}
G\omunu G\orhosigma(\nabla\umu\thetainv_{\rho\alpha})(\nabla\unu\thetainv_{\sigma\beta})g\oalphabeta
&=
G\omunu G\orhosigma(\partial\umu\thetainv_{\rho\alpha}-\Gamma\olambda_{\mu\rho}\thetainv_{\lambda\alpha}-
\Gamma\olambda_{\mu\alpha}\thetainv_{\rho\lambda})\times 
\\
&\quad
(\partial\unu\thetainv_{\sigma\beta}-\Gamma^{\lambda\oprime}_{\nu\sigma}\thetainv_{\lambda\oprime\beta}
-\Gamma^{\lambda\oprime}_{\nu\beta}\thetainv_{\sigma\lambda\oprime})g\oalphabeta \\
&=
G\omunu G\orhosigma(\partial\umu\thetainv_{\rho\alpha})(\partial\unu\thetainv_{\sigma\beta})
-2G\omunu G\orhosigma\Gamma\olambda_{\mu\rho}\thetainv_{\lambda\alpha}(\partial\unu\thetainv_{\sigma\beta})g\oalphabeta \\
&\quad
+2G\omunu \Gamma\olambda_{\mu\alpha}\theta^{\sigma\rho} g_{\rho\lambda}(\partial\unu\thetainv_{\sigma\beta})g\oalphabeta
+\Gamma\olambda_{\mu\rho}\Gamma^{\lambda\oprime}_{\nu\sigma}G\omunu G\orhosigma G_{\lambda\lambda\oprime} \\
&\quad
-2G\omunu \Gamma\olambda_{\mu\alpha}\Gamma^{\lambda\oprime}_{\nu\sigma}
\theta^{\sigma\rho}g_{\rho\lambda}\thetainv_{\lambda\oprime\beta}g\oalphabeta
+G\omunu \Gamma\olambda_{\mu\alpha}\Gamma^{\lambda\oprime}_{\nu\beta}g\oalphabeta g_{\lambda\lambda\oprime}
\end{split}
\end{align}
\begin{align}
\begin{split}
G\omunu G\orhosigma (\nabla\umu\thetainv_{\rho\alpha})(\nabla\usigma\thetainv_{\nu\beta})g\oalphabeta
&=
G\omunu G\orhosigma
(\partial\umu\thetainv_{\rho\alpha}-\Gamma\olambda_{\mu\rho}\thetainv_{\lambda\alpha}
-\Gamma\olambda_{\mu\alpha}\thetainv_{\rho\lambda})\times \\
&\quad
(\partial\usigma\thetainv_{\nu\beta}-\Gamma^{\lambda\oprime}_{\sigma\nu}\thetainv_{\lambda\oprime\beta}
-\Gamma^{\lambda\oprime}_{\sigma\beta}\thetainv_{\nu\lambda\oprime})g\oalphabeta \\
&=
G\omunu G\orhosigma (\partial\umu\thetainv_{\rho\alpha})(\partial\usigma\thetainv_{\nu\beta})g\oalphabeta
-2G\omunu G\orhosigma\Gamma\olambda_{\mu\rho}\thetainv_{\lambda\alpha}(\partial\usigma\thetainv_{\nu\beta})g\oalphabeta \\
&\quad
+2G\omunu \Gamma\olambda_{\mu\alpha}\theta^{\sigma\rho}g_{\rho\lambda}(\partial\usigma\thetainv_{\nu\beta})g\oalphabeta
+G\omunu G\orhosigma \Gamma\olambda_{\mu\rho}\Gamma^{\lambda\oprime}_{\sigma\nu}G_{\lambda\lambda\oprime}\\
&\quad
-2G\omunu\Gamma\olambda_{\mu\alpha}\Gamma^{\lambda\oprime}_{\sigma\nu}
\theta^{\sigma\rho}g_{\rho\lambda}\thetainv_{\lambda\oprime\beta}g\oalphabeta
+\Gamma\olambda_{\mu\alpha}\Gamma^{\lambda\oprime}_{\sigma\beta}\theta^{\sigma\rho}
g_{\rho\lambda}\theta^{\mu\nu}g_{\nu\lambda\oprime}g\oalphabeta
\end{split}
\end{align}
Combing the above three terms gives in total
\begin{align}
\begin{split}
&k\Big(G\omunu(\nabla\umu\thetainv_{\nu\alpha})G\orhosigma(\nabla\urho\thetainv_{\sigma\beta})g\oalphabeta 
-G\omunu G\orhosigma(\nabla\umu\thetainv_{\rho\alpha})(\nabla\unu\thetainv_{\sigma\beta})g\oalphabeta \\
&+G\omunu G\orhosigma (\nabla\umu\thetainv_{\rho\alpha})(\nabla\usigma\thetainv_{\nu\beta})g\oalphabeta\Big) =
k\Big(G\omunu(\partial\umu\thetainv_{\nu\alpha})G\orhosigma(\partial\urho\thetainv_{\sigma\beta})g\oalphabeta \\
&-G\omunu G\orhosigma(\partial\umu\thetainv_{\rho\alpha})(\partial\unu\thetainv_{\sigma\beta})g\oalphabeta 
+G\omunu G\orhosigma (\partial\umu\thetainv_{\rho\alpha})(\partial\usigma\thetainv_{\nu\beta})g\oalphabeta
\\
&-2G\omunu G\orhosigma\Gamma^{\lambda}\umunu\thetainv_{\lambda\alpha}(\partial\urho\thetainv_{\sigma\beta})g\oalphabeta 
+2\underbrace{\Gamma\olambda_{\mu\alpha}\theta\omunu g_{\nu\lambda}G\orhosigma(\partial\urho\thetainv_{\sigma\beta})g\oalphabeta}_{(c)}\\
&
+G\omunu G\orhosigma\Gamma\umunu\olambda\Gamma^{\lambda\oprime}\urhosigma G_{\lambda\lambda\oprime}
-2G\orhosigma\Gamma\olambda_{\mu\alpha}\Gamma^{\lambda\oprime}_{\rho\sigma}
\theta^{\mu\nu}g_{\nu\lambda}\thetainv_{\lambda\oprime\beta}g\oalphabeta\\
&
+\underbrace{\Gamma\olambda_{\mu\alpha}\Gamma^{\lambda\oprime}_{\rho\beta}\theta\omunu g_{\nu\lambda}
\theta\orhosigma g_{\sigma\lambda\oprime}g\oalphabeta}_{(d)}
+2G\omunu G\orhosigma\Gamma\olambda_{\mu\rho}\thetainv_{\lambda\alpha}(\partial\unu\thetainv_{\sigma\beta})g\oalphabeta \\
&
-2\underbrace{G\omunu \Gamma\olambda_{\mu\alpha}\theta^{\sigma\rho} g_{\rho\lambda}(\partial\unu\thetainv_{\sigma\beta})g\oalphabeta}_{(b)}
-\underbrace{G\omunu \Gamma\olambda_{\mu\alpha}\Gamma^{\lambda\oprime}_{\nu\beta}g\oalphabeta g_{\lambda\lambda\oprime}}_{(f)} \\
&
-2G\omunu G\orhosigma\Gamma\olambda_{\mu\rho}\thetainv_{\lambda\alpha}(\partial\usigma\thetainv_{\nu\beta})g\oalphabeta 
+2\underbrace{G\omunu \Gamma\olambda_{\mu\alpha}\theta^{\sigma\rho}g_{\rho\lambda}(\partial\usigma\thetainv_{\nu\beta})g\oalphabeta}_{(a)}\\
&
+\underbrace{\Gamma\olambda_{\mu\alpha}\Gamma^{\lambda\oprime}_{\sigma\beta}\theta^{\sigma\rho}
g_{\rho\lambda}\theta^{\mu\nu}g_{\nu\lambda\oprime}g\oalphabeta}_{(e)} \Big)
\\
&=k\Big(G\omunu(\partial\umu\thetainv_{\nu\alpha})G\orhosigma(\partial\urho\thetainv_{\sigma\beta})g\oalphabeta -G\omunu G\orhosigma(\partial\umu\thetainv_{\rho\alpha})(\partial\unu\thetainv_{\sigma\beta})g\oalphabeta \\
&
+G\omunu G\orhosigma (\partial\umu\thetainv_{\rho\alpha})(\partial\usigma\thetainv_{\nu\beta})g\oalphabeta
+\sum_{i=a}^f (i) \\
&
-2\,\eps G\omunu \Gamma\umunu\olambda G_{\lambda\eta}\widetilde{\Gamma}^{\eta}
+G\omunu G\orhosigma\Gamma\umunu\olambda\Gamma^{\eta}\urhosigma G_{\lambda\eta}
\Big)\\
&=
\tr a\omu a\onu G\umunu
-k\, G\omunu R\umunu[g] 
-k\,g\omunu \Gamma\umunu\olambda G\orhosigma \Gamma\urhosigma^\eta g_{\eta\lambda}\\
&
+k\,\eps g\omunu(\partial\umu\partial\unu \phi^i)
\left(
\Delta_{\Gtilde}\phi^j + \widetilde{\Gamma}\orho(\partial\urho \phi^j)
\right)\delta_{ij}\\
&
-2\,\eps G\omunu \Gamma\umunu\olambda G_{\lambda\eta}\widetilde{\Gamma}^{\eta}
+G\omunu G\orhosigma\Gamma\umunu\olambda\Gamma^{\eta}\urhosigma G_{\lambda\eta}
.
\end{split}
\end{align}
In the above equation two terms cancel due to
\begin{align}
\begin{split}
G\omunu G\orhosigma\Gamma\olambda_{\mu\rho}\thetainv_{\lambda\alpha}(\partial\unu\thetainv_{\sigma\beta})g\oalphabeta&=
G^{\rho\nu}G^{\mu\sigma}\Gamma\olambda_{\mu\rho}\thetainv_{\lambda\alpha}(\partial\usigma\thetainv_{\nu\beta})g\oalphabeta\\
&=
G\omunu G\orhosigma\Gamma\olambda_{\mu\rho}\thetainv_{\lambda\alpha}(\partial\usigma\thetainv_{\nu\beta})g\oalphabeta.
\end{split}
\end{align}
We also exploited the following relations, 
\begin{align}
\begin{split}
G\omunu G\orhosigma\Gamma^{\lambda}\umunu\thetainv_{\lambda\alpha}(\partial\urho\thetainv_{\sigma\beta})g\oalphabeta
+
G\orhosigma\Gamma\olambda_{\mu\alpha}\Gamma^{\lambda\oprime}_{\rho\sigma}
\theta^{\mu\nu}g_{\nu\lambda}\thetainv_{\lambda\oprime\beta}g\oalphabeta
&= 
G\omunu \Gamma\umunu\olambda \thetainv_{\lambda\beta}g\oalphabeta \times \\
&\quad
\left(
G\orhosigma (\partial\urho \thetainv_{\sigma\alpha})
+\theta\orhosigma \Gamma^{\eta}_{\rho\alpha}g_{\sigma\eta}
\right) \\
&= 
-G\omunu \Gamma\umunu\olambda G_{\lambda\sigma} \times \\
&\quad
\left(
\theta^{\rho\alpha}(\partial\urho\theta^{\sigma\beta})g\ualphabeta
+\theta^{\rho\alpha}\theta^{\sigma\beta}(\partial\urho g\ualphabeta)
\right) \\
&=
\eps G\omunu \Gamma\umunu\olambda G_{\lambda\sigma}\widetilde{\Gamma}\osigma,
\end{split}
\end{align}
and
\begin{align}
\begin{split}
\theta^{\rho\sigma}(\partial\urho g_{\sigma\beta})=
\Gamma^{\lambda}_{\rho\alpha}g_{\lambda\sigma}\theta\orhosigma.
\end{split}
\end{align}

In the end we obtain for $\tr a\omu a\onu G\umunu$  the following result, 
\begin{align}
\begin{split}
\tr\,a\omu a\onu G\umunu 
&=
k\Big(
G\omunu(\nabla\umu\thetainv_{\nu\alpha})G\orhosigma(\nabla\urho\thetainv_{\sigma\beta})g\oalphabeta 
-G\omunu G\orhosigma(\nabla\umu\thetainv_{\rho\alpha})(\nabla\unu\thetainv_{\sigma\beta})g\oalphabeta \\
&\quad
+G\omunu G\orhosigma (\nabla\umu\thetainv_{\rho\alpha})(\nabla\usigma\thetainv_{\nu\beta})g\oalphabeta 
+  G\omunu R\umunu[g] 
+  g\omunu \Gamma\olambda\umunu G\orhosigma \Gamma^{\eta}\urhosigma g_{\eta\lambda} \\
&\quad
-  \eps g\omunu (\partial\umu\partial\unu\phi^i)
\left(\Delta_{\Gtilde}\phi^j +\widetilde{\Gamma}\orho(\partial\urho\phi^j)\right)\delta_{ij}
+2  G\omunu \Gamma\olambda\umunu \widetilde{\Gamma}^{\eta}\Gtilde_{\lambda\eta} \\
&\quad
-G\omunu G\orhosigma \Gamma\olambda\umunu \Gamma^{\eta}\urhosigma G_{\lambda\eta}
\Big).
\end{split}
\end{align}

\paragraph{General case: $\Gtilde \neq g$ but using e.o.m.}
We have
\begin{align}
\begin{split}
\eps G\omunu \Gamma\olambda\umunu \widetilde{\Gamma}^{\eta}G_{\lambda\eta}=G\omunu G\orhosigma \Gamma\olambda\umunu \Gamma^{\lambda\oprime}\urhosigma G_{\lambda\lambda\oprime}
= G\orhosigma g^{\delta\beta} \Gamma^{\eta}_{\delta\beta} \Gamma^{\lambda}\urhosigma g_{\eta\lambda}=0.
\end{split}
\end{align}
This can be seen by considering on-shell configurations $\Delta_{\Gtilde}\phi^i=0$, $\widetilde{\Gamma}\omu=0$ which imply  
\begin{align}
\begin{split}
G\omunu \Gamma\olambda\umunu &= G\omunu (\partial\urho\phi^i)(\partial\umu\partial\unu\phi^j)g^{\rho\lambda}\delta_{ij}\\
&=\eps  g^{\rho\lambda}(\partial\urho\phi^i)\Gtilde\omunu(\partial\umu\partial\unu\phi^j)\delta_{ij} \\
&= 0.
\end{split}
\end{align}

We have shown that for on-shell geometries $\tr\cE$ is indeed a covariant expression,
\begin{align}\label{eq: covariant E II}
\begin{split}
\tr \cE &=
- \frac{\ems}{4}
\tr\, G\umunu a\omu a\onu \\
 &= -\frac{\ems}{4}\,k\,
\Big(
G\omunu(\nabla\umu\thetainv_{\nu\alpha})G\orhosigma(\nabla\urho\thetainv_{\sigma\beta})g\oalphabeta 
-G\omunu G\orhosigma(\nabla\umu\thetainv_{\rho\alpha})(\nabla\unu\thetainv_{\sigma\beta})g\oalphabeta \\
&\quad 
+G\omunu G\orhosigma (\nabla\umu\thetainv_{\rho\alpha})(\nabla\usigma\thetainv_{\nu\beta})g\oalphabeta
\Big)
-\frac{k}{4}\Gtilde\omunu R\umunu[g].
\end{split}
\end{align}
Hence we can go to normal coordinates to simplify the comparison between $\tr\cE$ and the Ricci scalar $R[\Gtilde]$. Keep in mind that the covariant derivative in Eq. (\ref{eq: covariant E II}) is with respect to the background metric $g\umunu$.

\paragraph{Special case: $\Gtilde=g$ without the use of e.o.m.}

In that case we have 
\begin{align}
\begin{split}
G\omunu\Gamma\olambda\umunu \widetilde{\Gamma}^{\eta} \Gtilde_{\lambda\eta}
=
G\omunu G\orhosigma \Gamma\olambda\umunu \Gamma^{\eta}\urhosigma G_{\lambda\eta}
=
g\omunu \Gamma\olambda\umunu G\orhosigma \Gamma^{\eta}\urhosigma g_{\lambda\eta}
=\eps \Gamma\olambda \Gamma^{\eta} g_{\lambda\eta}.
\end{split}
\end{align}
So we find for $\tr\cE$
\begin{align}
\begin{split}
\tr \cE 
 &=
 -\frac{\ems}{4}\tr(a\omu a\onu G\umunu) +\frac{k}{4} \Gamma\olambda \Gamma^{\eta} g_{\lambda\eta} \\ 
 &=
 -\frac{\eps}{4}\,k\,
\Big(
g\omunu(\nabla\umu\thetainv_{\nu\alpha})g\orhosigma(\nabla\urho\thetainv_{\sigma\beta})g\oalphabeta 
-g\omunu g\orhosigma(\nabla\umu\thetainv_{\rho\alpha})(\nabla\unu\thetainv_{\sigma\beta})g\oalphabeta \\
&\quad 
+g\omunu g\orhosigma (\nabla\umu\thetainv_{\rho\alpha})(\nabla\usigma\thetainv_{\nu\beta})g\oalphabeta
\Big)
-\frac{k}{4} g\omunu R\umunu[g] -\frac{1}{2}\Gamma\omu \Gamma\onu g\umunu\\
&\quad
+\frac{\ems}{4} k
\left( \Delta_g \phi^i + \Gamma\omu (\partial\umu \phi^i) \right)
\left( \Delta_g \phi^j + \Gamma\onu (\partial\unu \phi^j) \right)\delta_{ij}\\
&\quad
+  \frac{1}{4}k\, g\umunu \Gamma\omu\Gamma\onu.
\end{split}
\end{align}

\begin{align}
\begin{split}
\Gamma\omu &= g\omunu g\orhosigma (\partial\urho g_{\sigma\nu})
-\frac{1}{2}g\omunu g\orhosigma \partial\unu g\urhosigma \\
&=
g\omunu g\orhosigma (\partial\urho\partial\usigma\phi^i)(\partial\unu \phi^j)\delta_{ij} \\
&=
g\omunu (\partial\unu\phi^j) \delta_{ij}\left(
\Delta_g \phi^i + \Gamma\orho (\partial\urho \phi^i)
\right)
\end{split}
\end{align}
or
\begin{align}
g_{\rho\mu}\Gamma\omu =
(\partial\urho\phi^i)\left( 
\Delta_g \phi^j + \Gamma\omu (\partial\umu \phi^j)
\right)\delta_{ij}.
\end{align}
Recalling  
\begin{align}
\begin{split}
\Delta_g x^a = \left( \begin{array}{c} \Delta_g x\omu \\ \Delta_g \phi^i \end{array}\right)
=
\left( \begin{array}{c} -\Gamma\omu \\ \Delta_g \phi^i \end{array}\right)
\end{split}
\end{align}
we see that
\begin{align}
\begin{split}
\partial\urho x^a \Delta_g x^b \eta_{ab}= -\Gamma\omu \eta_{\mu\rho}+\partial\urho \phi^i \Delta_g \phi^j \delta_{ij},
\end{split}
\end{align}
as well as
\begin{align}
\begin{split}
g_{\rho\mu}\Gamma\omu &=
(\partial\urho\phi^i)(\Delta_g\phi^j)\delta_{ij} + \Gamma\omu (\partial\umu\phi^i)(\partial\urho\phi^j)\delta_{ij} \\
&=
(\partial\urho x^a )(\Delta_g x^b)\eta_{ab}+\Gamma\umu \eta_{\mu\rho}
+ \Gamma\omu (\partial\umu\phi^i)(\partial\urho\phi^j)\delta_{ij} \\
&= (\partial\urho x^a )(\Delta_g x^b)\eta_{ab} + \Gamma\omu g_{\mu\rho}.
\end{split}
\end{align}
Therefore we have
\begin{align}\label{eq: eom 3}
(\partial\urho x^a)(\Delta_g x ^b )\eta_{ab}=0
\end{align}
or
\begin{align}
(\partial\urho\phi^i)(\Delta_g\phi^j)\delta_{ij}=\Gamma\omu \eta_{\mu\rho}.
\end{align}
It is worthwhile mentioning that the relation Eq. (\ref{eq: eom 3}) for the general case $\Gtilde \neq g$ 
turns out to be an e.o.m~\cite{Steinacker:2008ya} and thus this equation usually holds only for on-shell geometries. 
\begin{align}
\begin{split}
\left( \Delta_g \phi^i +\Gamma\omu(\partial\umu\phi^i) \right)\left( \Delta_g \phi^j +\Gamma\onu(\partial\unu\phi^j) \right)\delta_{ij} &= 
\Big((\Delta_g\phi^i)(\Delta_g\phi^j) 
+2\Gamma\omu(\partial\umu\phi^i)(\Delta_g\phi^j)\\
&\quad
+\Gamma\omu \Gamma\onu (\partial\umu\phi^i)(\partial\unu\phi^j)\Big)\delta_{ij}  \\
&=
(\Delta_g\phi^i)(\Delta_g\phi^j)\delta_{ij}
+2\Gamma\onu \eta\umunu\\
&\quad
+\Gamma\omu \Gamma\onu (\partial\umu\phi^i)(\partial\unu\phi^j)\delta_{ij}  \\
&=
(\Delta_g\phi^i)(\Delta_g\phi^j)\delta_{ij}
+\Gamma\omu \Gamma\onu \eta\umunu \\
&\quad
+ \Gamma\omu \Gamma\onu g\umunu  \\
&=
(\Delta_g\phi^i)(\Delta_g\phi^j)\delta_{ij}
+(\Delta_g x\omu)(\Delta_g x\onu)\eta\umunu \\
&\quad + \Gamma\omu \Gamma\onu g\umunu  \\
&=
(\Delta_g x^a)(\Delta_g x^b)\eta_{ab}+  \Gamma\omu \Gamma\onu g\umunu 
\end{split}
\end{align}

\begin{align}
\begin{split}
\tr \cE 
 &=
 -\frac{\ems}{4}\tr(a\omu a\onu G\umunu) +\frac{k}{4} \Gamma\olambda \Gamma^{\eta} g_{\lambda\eta} \\ 
 &=
 -\frac{\eps}{4}\,k\,
\Big(
g\omunu(\nabla\umu\thetainv_{\nu\alpha})g\orhosigma(\nabla\urho\thetainv_{\sigma\beta})g\oalphabeta 
-g\omunu g\orhosigma(\nabla\umu\thetainv_{\rho\alpha})(\nabla\unu\thetainv_{\sigma\beta})g\oalphabeta \\
&\quad 
+g\omunu g\orhosigma (\nabla\umu\thetainv_{\rho\alpha})(\nabla\usigma\thetainv_{\nu\beta})g\oalphabeta
\Big)
-\frac{k}{4} R[g] -\frac{1}{2}\Gamma\omu \Gamma\onu g\umunu\\
&\quad
+\frac{k}{4}(\Delta_g x^a)(\Delta_b x^b)\eta_{ab} + \frac{k}{4}\,\Gamma\omu\Gamma\onu g\umunu
+  \frac{k}{4} \Gamma\omu\Gamma\onu g\umunu \\
&=
-\frac{\eps}{4}\,k\,
\Big(
g\omunu(\nabla\umu\thetainv_{\nu\alpha})g\orhosigma(\nabla\urho\thetainv_{\sigma\beta})g\oalphabeta 
-g\omunu g\orhosigma(\nabla\umu\thetainv_{\rho\alpha})(\nabla\unu\thetainv_{\sigma\beta})g\oalphabeta \\
&\quad 
+g\omunu g\orhosigma (\nabla\umu\thetainv_{\rho\alpha})(\nabla\usigma\thetainv_{\nu\beta})g\oalphabeta
\Big)
-\frac{k}{4} R[g]+\frac{k}{4}(\Delta_g x^a)(\Delta_b x^b)\eta_{ab}.
\end{split}
\end{align}

Due to the antisymmetry of $\theta\omunu$ and since $\theta\omunu$ also fulfills the Jacobi identity we have
\begin{align}
\begin{split}
\nabla\urho\thetainv\umunu + \nabla\unu\thetainv_{\rho\mu}+\nabla\umu\thetainv_{\nu\rho}=0
\end{split}
\end{align}
Via the following computation 
\begin{align}
\begin{split}
G\omunu G\orhosigma (\nabla\umu\thetainv_{\rho\alpha})(\nabla\unu\thetainv_{\sigma\beta})g\oalphabeta
&=
G\omunu G\orhosigma 
\left( \nabla\ualpha\thetainv_{\mu\rho}+\nabla\urho \thetainv_{\alpha\mu}\right)
\left( \nabla\ubeta\thetainv_{\nu\sigma}+\nabla\usigma\thetainv_{\beta\nu}\right)g\oalphabeta \\
&=
G\omunu G\orhosigma
\Big((\nabla\ualpha\thetainv_{\mu\rho})(\nabla\ubeta\thetainv_{\nu\sigma})
+2(\nabla\urho\thetainv_{\alpha\mu})(\nabla\ubeta\thetainv_{\nu\sigma})\\
&\quad
+(\nabla\umu\thetainv_{\rho\alpha})(\nabla\unu\thetainv_{\sigma\beta})
\Big)g\oalphabeta
\end{split}
\end{align}
we see that
\begin{align}
G\omunu G\orhosigma
\Big((\nabla\ualpha\thetainv_{\mu\rho})(\nabla\ubeta\thetainv_{\nu\sigma})
+2(\nabla\urho\thetainv_{\alpha\mu})(\nabla\ubeta\thetainv_{\nu\sigma})\Big)g\oalphabeta=0
\end{align}
In the case of $\Gtilde\umunu=g\umunu$ we hence have
\begin{align}
g\omunu g\orhosigma g\oalphabeta (\nabla\umu \thetainv_{\rho\alpha})(\nabla\unu\thetainv_{\sigma\beta})=
2g\omunu g\orhosigma g\oalphabeta(\nabla\umu
\thetainv_{\rho\alpha})(\nabla\usigma\thetainv_{\nu\beta}).
\label{bianci}
\end{align}
Also, in that case the following relation holds
\begin{align}\label{eq: homogeneous maxwell}
g\omunu \nabla\umu \thetainv_{\nu\rho}=0.
\end{align}
To see this consider the covariant derivative acting on the Poisson structure
\begin{align}
\begin{split}
g\omunu \nabla\umu \thetainv_{\nu\rho}&=
g\omunu \partial\umu \thetainv_{\nu\rho}-g\omunu\Gamma\olambda\umunu \thetainv_{\lambda\rho}
-g\omunu \Gamma\olambda_{\mu\rho}\thetainv_{\nu\lambda} .
\end{split}
\end{align}
Using
\begin{align}
\begin{split}
g\omunu \Gamma\olambda_{\mu\rho}\thetainv_{\lambda\nu}
&=\Big(
\frac{1}{2}g\omunu g^{\lambda\eta}(\partial\umu g_{\eta \rho})
-\frac{1}{2}g\omunu g^{\lambda\eta}(\partial_{\eta}g_{\mu\rho})
-\frac{1}{2}(\partial\urho g^{\lambda\nu})
\Big)\thetainv_{\nu\lambda} \\
&=
g\omunu g^{\lambda\eta}(\partial\umu g_{\eta\rho})\thetainv_{\nu\lambda}\\
&=-\ems \theta^{\mu\eta}(\partial\umu g_{\eta\rho})
\end{split}
\end{align}
we can see that
\begin{align}
\begin{split}
g\omunu \nabla\umu \thetainv_{\nu\rho}&=g\omunu(\partial\umu \thetainv_{\nu\rho})
+\ems \theta^{\mu\eta}(\partial\umu g_{\nu\rho}) -\Gamma\olambda \thetainv_{\lambda\rho} \\
&=
g\omunu(\partial\umu \thetainv_{\nu\rho})
+\ems \theta^{\mu\eta}(\partial\umu g_{\nu\rho}) \\
&\quad
+\ems(\partial_{\eta}\theta^{\lambda\alpha})\theta^{\eta\beta}g\ualphabeta \thetainv_{\lambda\rho}
+\ems\theta^{\lambda\alpha}\theta^{\eta\beta}(\partial\urho g\ualphabeta)\thetainv_{\lambda\rho} \\
&=0
\end{split}
\end{align}

The covariant e.o.m. that was derived in~\cite{Steinacker:2008ri}
\begin{align}
\Gtilde\omunu\widetilde{\nabla}\umu (\eps \thetainv_{\nu\rho})
=
\frac{\ems}{4} \Gtilde_{\rho\mu}\theta\omunu \partial\unu(G^{\kappa\lambda}g_{\kappa\lambda})
\end{align}
reduces for $\Gtilde\umunu = g\umunu$ to
\begin{align}
g\omunu \nabla\umu \thetainv_{\nu\rho}=0,
\end{align}
which has the form of a homogeneous Maxwell equation. So for $\Gtilde = g$ this relation is actually an identity. 

Our final result for $\tr \cE$ in case of $\Gtilde\umunu = g\umunu$ is 
\begin{align}\label{eq: covariant E III}
\begin{split}
\tr \cE 
 &= 
 \frac{\eps}{4}\,k\,
g\omunu g\orhosigma (\nabla\umu\thetainv_{\rho\alpha})(\nabla\usigma\thetainv_{\nu\beta})g\oalphabeta 
-\frac{k}{4} R[g] 
+\frac{k}{4}  (\Delta_g x^a)(\Delta_g x^b)\eta_{ab}.
\end{split}
\end{align}
Thus we have shown that in the case of $\Gtilde\umunu = g\umunu$, 
for a covariance proof it is not necessary to use e.o.m.

\paragraph{Special case $\Gtilde = g$ using e.o.m}

$\tr\cE$ is now very simple,
\begin{align}
\begin{split}
\tr \cE
 &= \frac{\eps}{4}\,k\,
 g\omunu g\orhosigma (\nabla\umu\thetainv_{\rho\alpha})(\nabla\usigma\thetainv_{\nu\beta})g\oalphabeta 
-\frac{k}{4} R[g] .
\end{split}
\end{align}

\section{Appendix C: Expressing $R$ in normal coordinates}
We evaluate the Ricci scalar in normal coordinates. First of all note that
\begin{align}
\Gtilde\orhosigma\partial\umu \Gtilde\urhosigma = g\orhosigma\partial\umu g\urhosigma
\stackrel{nc}{=}0.
\end{align}
Using also
\begin{align}
\Gtilde\omunu \Gtilde\orhosigma (\partial\umu\partial\urho \Gtilde_{\nu\sigma})
= -\Gtilde\omunu(\partial\umu \Gtilde\orhosigma)(\partial\urho \Gtilde_{\sigma\nu})
+\Gtilde\umunu(\partial\urho \Gtilde^{\rho\mu})(\partial\usigma \Gtilde^{\sigma\nu})
-\partial\umu \partial\unu \Gtilde\omunu
\end{align}
we can simplify the Ricci scalar Eq.(\ref{eq: R}) and we obtain
\begin{align}
\begin{split}
R[\Gtilde]
&=
\ems\Big\{
-\frac{3}{2}G\omunu(\partial\umu\sigma)(\partial\unu\sigma)
-3G\omunu(\partial\umu\partial\unu\sigma)\\
&\quad
+(\partial\umu G\omunu)(\partial\unu\sigma) 
-\frac{1}{2}G\omunu(\partial\umu\sigma)(G\orhosigma\partial\unu G\urhosigma) \\
&\quad
-\frac{1}{2}G\omunu(\partial\umu G\orhosigma)(\partial\urho G_{\sigma\nu})
-\partial\umu\partial\unu G\omunu \\
&\quad
-G\omunu(G\orhosigma\partial\umu\partial\unu G\urhosigma)
-\frac{3}{4}G\omunu(\partial\umu G\orhosigma)(\partial\unu G\urhosigma)
\Big\}.
\end{split}
\end{align}
Next give a list of terms that appear in the Ricci scalar in normal coordinates. Also here we exploit the e.o.m. Eq. (\ref{eom}). 
\begin{align}
\begin{split}
G\omunu(\partial\umu\sigma)(\partial\unu\sigma)&=(\partial\umu\theta^{\mu\alpha})(\partial\unu\theta^{\nu\beta})g\ualphabeta \\
G\omunu(\partial\umu\partial\unu \sigma)&=\frac{1}{2}G\omunu(\theta\orhosigma\partial\umu\partial\unu\theta\urhosigma)
+\frac{1}{2}G\omunu(\partial\umu\theta\orhosigma)(\partial\unu\theta\urhosigma)\\
&\quad
+\frac{1}{2}G\omunu(g\orhosigma\partial\umu \partial\unu g\urhosigma) \\
(\partial\umu G\omunu)(\partial\unu\sigma)
&\stackrel{eom}{=}(\partial\umu\theta^{\mu\alpha})(\partial\unu\theta^{\nu\beta})g\ualphabeta  \\
(G\orhosigma\partial\umu G\urhosigma)&=-4(\partial\umu\sigma)\\
G\omunu(\partial\umu\sigma)(G\orhosigma\partial\unu G\urhosigma)
&=-4(\partial\umu\theta^{\mu\alpha})(\partial\unu\theta^{\nu\beta})g\ualphabeta 
\end{split}
\end{align}
\begin{align}
\begin{split}
G\omunu(\partial\umu G\orhosigma)(\partial\unu G\urhosigma)&=
-2G\omunu(\partial\umu\theta\orhosigma)(\partial\unu\theta\urhosigma)
-2G\omunu G\orhosigma(\partial\umu\thetainv_{\rho\alpha})(\partial\unu\thetainv_{\sigma\beta})g\oalphabeta\\
(\partial\umu G\orhosigma)(\partial\unu G\urhosigma)+ G\orhosigma\partial\umu\partial\unu G\urhosigma
&=
-4\partial\umu\partial\unu \sigma +g\orhosigma(\partial\umu\partial\unu g\urhosigma) \\
G\omunu(G\orhosigma\partial\umu\partial\unu G\urhosigma)&=
-G\omunu(\partial\umu G\orhosigma)(\partial\unu G\urhosigma)-4\partial\umu\partial\unu \sigma
+G\omunu(g\orhosigma\partial\umu\partial\unu g\urhosigma) \\
&=2 G\omunu G\orhosigma(\partial\umu\thetainv_{\rho\alpha})(\partial\unu \thetainv_{\sigma\beta})g\oalphabeta
-2G\omunu(\theta\orhosigma\partial\umu\partial\unu\theta\urhosigma)\\
&\quad
- G\omunu(g\orhosigma\partial\umu\partial\unu g\urhosigma)
\end{split}
\end{align}
\begin{align}
\begin{split}
G\omunu(\partial\umu G\orhosigma)(\partial\urho G_{\sigma\nu})&=
-(\partial\umu\theta^{\nu\alpha})(\partial\unu \theta^{\mu\beta})g\ualphabeta
-2G\omunu (\partial\umu\theta^{\rho\alpha})(\partial\urho \thetainv_{\nu\alpha})\\
&\quad
-G\omunu G\orhosigma(\partial\umu\thetainv_{\rho\alpha})(\partial\usigma\thetainv_{\nu\beta})g\oalphabeta \\
\partial\umu\partial\unu G\omunu 
&\stackrel{eom}{=}
\theta^{\mu\alpha}(\partial\umu\partial\unu\theta^{\nu\beta})g\ualphabeta
+(\partial\umu\theta^{\mu\alpha})(\partial\unu\theta^{\nu\beta})g\ualphabeta \\
&=
\frac{1}{2}G\omunu(\theta\orhosigma\partial\umu\partial\unu\theta\urhosigma)
+\frac{1}{2}G\omunu(\partial\umu\theta\orhosigma)(\partial\unu\theta\urhosigma)\\
&\quad
+(\partial\umu\theta^{\mu\alpha})(\partial\unu\theta^{\nu\beta})g\ualphabeta
\end{split}
\end{align}
Using these we get our final result
\begin{align}
\begin{split}
R[\Gtilde]\stackrel{nc}{=}&\ems\Big\{
\frac{1}{2}(\partial\umu\theta^{\mu\alpha})(\partial\unu\theta^{\nu\beta})\eta\ualphabeta
+\frac{1}{2}(\partial\umu\theta^{\nu\alpha})(\partial\unu\theta^{\mu\beta})\eta\ualphabeta \\
&
+\frac{1}{2}G\omunu G\orhosigma(\partial\umu\thetainv_{\rho\alpha})(\partial\usigma\thetainv_{\nu\beta})\eta\oalphabeta
-\frac{1}{2}G\omunu G\orhosigma(\partial\umu\thetainv_{\rho\alpha})(\partial\unu\thetainv_{\sigma\beta})\eta\oalphabeta \\
&
-\frac{1}{2}G\omunu(g\orhosigma\partial\umu \partial\unu g\urhosigma)
\Big\}.
\end{split}
\end{align}


\begin{thebibliography}{99}

\bibitem{Szabo:2001kg}
     R.~Szabo, 
     ``Quantum field theory on noncommutative spaces,''
     Phys.\ Rept.\ {\bf 378},
    (2003) 207, [arXiv:hep-th/0109162].



\bibitem{Steinacker:2007dq}
  H.~Steinacker,
  ``Emergent Gravity from Noncommutative Gauge Theory,''
  JHEP {\bf 12}, (2007) 049; 
 [arXiv:0708.2426 [hep-th]].

\bibitem{Grosse:2008xr}
  H.~Grosse, H.~Steinacker and M.~Wohlgenannt,
  ``Emergent Gravity, Matrix Models and UV/IR Mixing,''
  JHEP {\bf 0804} (2008) 023
  [arXiv:0802.0973 [hep-th]].


\bibitem{Klammer:2008df}
D.~Klammer and H.~Steinacker,
  ``Fermions and Emergent Noncommutative Gravity,''
  JHEP {\bf 0808} (2008) 074
  [arXiv:0805.1157 [hep-th]].

\bibitem{Steinacker:2008ri}
  H.~Steinacker,
  ``Emergent Gravity and Noncommutative Branes from Yang-Mills Matrix Models,''
  Nucl.\ Phys. {\bf B 810} (2009) 01;
  [arXiv:0806.2032 [hep-th]].

\bibitem{Steinacker:2008ya}
  H.~Steinacker,
  ``Covariant Field Equations, Gauge Fields and Conservation Laws from
  Yang-Mills Matrix Models,''
  JHEP {\bf 0902} (2009) 044
  [arXiv:0812.3761 [hep-th]].
  
\bibitem{Klammer:2009ku}
  D.~Klammer and H.~Steinacker,
  ``Cosmological solutions of emergent noncommutative gravity,''
  Phys.\ Rev.\ Lett.\  {\bf 102} (2009) 221301
  [arXiv:0903.0986 [gr-qc]].
  
\bibitem{Steinacker:2009mp}
     H.~Steinacker, ``On the Newtonian limit of emergent NC gravity and long-
                  distance corrections,'' (2009)
                  [arXiv:0909.4621 [hep-th]].  
                  
                  
\bibitem{Gilkey:1995mj}
  P.~B.~Gilkey,
  ``Invariance theory, the heat equation and the 
  Atiyah-Singer index theorem,''
  Wilmington, Publish or Perish, 1984.


\bibitem{Vassilevich:2003xt}
     D.V.~Vassilevich,
     ``Heat kernel expansion: User's manual'',
     Phys. \ Rept. {\bf 388} (2003) 279 
     [arXiv:hep-th/0306138].

\bibitem{Ishibashi:1996xs}
  N.~Ishibashi, H.~Kawai, Y.~Kitazawa and A.~Tsuchiya,
  ``A large-N reduced model as superstring,''
  Nucl.\ Phys.\  B {\bf 498} (1997) 467
  [arXiv:hep-th/9612115].
  

\bibitem{Kontsevich:1997vb}
     M.~Kontsevich,
     ``Deformation quantization of Poisson manifolds, I,''
     Lett.\ Math.\ Phys.\ {\bf 66} (2003) 157,
     [arXiv:q-alg/9709040].
      
\bibitem{weinberg}
S. Weinberg, ``Gravitation and Cosmology'' New York: Wiley.

\bibitem{Sakharov:1967pk}
     A. D.~Sakharov, ``Vacuum quantum fluctuations in curved space and the theory
                  of gravitation,''
     Sov.\ Phys.\ Dokl.\ {\bf 12} (1968) 1040.



\end{thebibliography}

\end{document}